\documentclass[twocolumn,twoside,11pt]{article}
\usepackage[left=0.58in,top=0.72in,bottom=0.7in,right=0.58in,headsep=0.28in,footskip=0.3in,footnotesep=4pt]{geometry}
\usepackage[para]{footmisc}
\usepackage[misc]{ifsym}
\usepackage{graphicx}
\graphicspath{{Figures/}}
\usepackage{dcolumn}
\usepackage{xcolor}
\usepackage{fancyhdr}
\usepackage{lipsum}
\usepackage{soul}
\usepackage{braket}
\usepackage{amsmath,amssymb}
\usepackage{titlesec}
\usepackage[colorlinks=true,citecolor=blue,linkcolor=blue,urlcolor=blue]{hyperref}
\usepackage{lmodern}
\usepackage{booktabs}
\usepackage{footnote}
\usepackage{lettrine}
\usepackage{cite}
\makeatletter 
\renewcommand{\fnum@figure}{\textbf{Fig.~\thefigure}}
\makeatother
\makeatletter
\def\bbordermatrix#1{\begingroup \m@th
  \@tempdima 4.75\p@
  \setbox\z@\vbox{%
    \def\cr{\crcr\noalign{\kern2\p@\global\let\cr\endline}}%
    \ialign{$##$\hfil\kern2\p@\kern\@tempdima&\thinspace\hfil$##$\hfil
      &&\quad\hfil$##$\hfil\crcr
      \omit\strut\hfil\crcr\noalign{\kern-\baselineskip}%
      #1\crcr\omit\strut\cr}}%
  \setbox\tw@\vbox{\unvcopy\z@\global\setbox\@ne\lastbox}%
  \setbox\tw@\hbox{\unhbox\@ne\unskip\global\setbox\@ne\lastbox}%
  \setbox\tw@\hbox{$\kern\wd\@ne\kern-\@tempdima\left[\kern-\wd\@ne
    \global\setbox\@ne\vbox{\box\@ne\kern2\p@}%
    \vcenter{\kern-\ht\@ne\unvbox\z@\kern-\baselineskip}\,\right]$}%
  \null\;\vbox{\kern\ht\@ne\box\tw@}\endgroup}
\makeatother

\usepackage{array}
\newcolumntype{L}{>{\centering\arraybackslash}m{2 cm}}
\newcolumntype{C}{>{\centering\arraybackslash}m{3.2 cm}}
\usepackage{multirow}
\usepackage{changepage}
\newcommand{\periodafter}[1]{#1.}
\definecolor{customgreen}{rgb}{0.306,0.463,0.51}
\titleformat{\section}[display]
  {\sffamily\large\bfseries}{}{0pt}{\large}
\titlespacing*{\section}{0pt}{0pt}{2pt}
\titleformat{\subsection}[runin]
  {\normalfont\bfseries}{}{0pt}{\periodafter}

\makeatletter
\def\@fnsymbol#1{\ensuremath{\ifcase#1\or *\or \dagger\or dagger\or \mathsection\or \mathparagraph\or \|\or **\or \dagger\dagger \or \text{\Letter} \else\@ctrerr\fi}}
\makeatother
\newcommand{\corresauth}[2][9]{\renewcommand{\thefootnote}{\fnsymbol{footnote}}\footnote[#1]{#2}\renewcommand{\thefootnote}{\arabic{footnote}}}

\usepackage{array}
\newcolumntype{L}{>{\centering\arraybackslash}m{2 cm}}
\newcolumntype{C}{>{\centering\arraybackslash}m{3.2 cm}}
\usepackage{multirow}

\begin{document}

\onecolumn
\renewenvironment{abstract}
{\begin{adjustwidth}{0pt}{133.7pt} \sffamily}
{\end{adjustwidth}}

\begin{titlepage}
\pagestyle{fancy}
\null\par
\vskip 9em
\setlength{\parindent}{0pt}
{\sffamily\Huge Accelerated Quantum Monte Carlo with Probabilistic Computers \par}
\vspace{2em}
{\sffamily Shuvro Chowdhury{\vspace{2em}} \footnote[1]{\sffamily Department of Electrical and Computer Engineering, University of California, Santa Barbara, Santa Barbara, CA 93106, USA,}\textsuperscript{,}\corresauth[9]{\sffamily email:  schowdhury@ucsb.edu}, Kerem Y. Camsari \footnotemark[1] and Supriyo Datta \footnote[2]{Elmore Family School of Electrical and Computer Engineering, Purdue University, West Lafayette, IN 47907, USA}}

\vspace{0.25in}

\begin{abstract}
Abstract -- Quantum Monte Carlo (QMC) techniques are widely used in a variety of scientific problems and much work has been dedicated to developing optimized algorithms that can accelerate QMC on standard processors (CPU). With the advent of various special purpose devices and domain specific hardware, it has become increasingly important to establish clear benchmarks of what improvements these technologies offer compared to existing technologies. In this paper, we demonstrate $2 \text{ to } 3$ orders of magnitude acceleration of a standard QMC algorithm using a specially designed digital processor, and a further $2 \text{ to } 3$ orders of magnitude by mapping it to a clockless analog processor. Our demonstration provides a roadmap for $5 \text{ to } 6$ orders of magnitude acceleration for a transverse field Ising model (TFIM) and could possibly be extended to other QMC models as well. The clockless analog hardware can be viewed as the classical counterpart of the quantum annealer and provides performance within a factor of $<10$ of the latter. The convergence time for the clockless analog hardware scales with the number of qubits as $\sim N$, improving the $\sim N^2$ scaling for CPU implementations, but appears worse than that reported for quantum annealers by D-Wave.
\end{abstract}

\vfill

{\color{customgreen}\rule{\textwidth}{.3ex}}
\end{titlepage}
\newpage
\phantomsection
\twocolumn
\section{Introduction}
\label{sec:Intro}

Envisioned by Feynman \cite{feynman1982simulating} and later formalized by Deutsch \cite{Deutsch1985} and others \cite{bernstein1993quantum}, quantum computing has been perceived by many as the natural simulator of quantum mechanical processes that govern natural phenomena. It became more popular with the discovery of powerful algorithms like Shor's integer factorization \cite{Shor1999} and Grover's search \cite{Grover96afast} offering significant theoretical speedup over their classical counterpart. A different flavor of quantum computing was also theorized in \cite{Farhi2000,Farhi2001,Reichardt2004,Smelyanskiy2001} which makes use of the adiabatic theorem \cite{Born1928}. It was later shown that these two flavors of quantum computing are equivalent \cite{Aharonov2007}. The technological difficulties of realizing noiseless qubits with coherent interactions among the qubits have focused recent efforts on the Noisy Intermediate Scale Quantum (NISQ) regime \cite{Preskill2018} and serious progress has been made in recent years \cite{Arute2019,Zhong1460,Neill2021,Huggins2022,Wu_2021,Brod2022,Madsen2022}. 

In the absence of general-purpose quantum computers, quantum Monte Carlo (QMC) still remains the standard tool to understand quantum many-body systems and to investigate a wide range of quantum phenomena -- including magnetic phase transitions, molecular dynamics, and astrophysics \cite{austin2012,Tews2020,Carlson2015,LOMNITZADLER1981399}. Much effort has been made to develop efficient QMC algorithms of various sorts \cite{Carlson2015,Blankenbecler1981,LOMNITZADLER1981399,Evertz1993,Bertrand2019,Cohen2015,VanHoucke2012,Bour2015,VANHOUCKE201095,LEE2009117} which can be suitably implemented on standard general-purpose classical processors (CPU). Interestingly for many important quantum problems, the efficiency of QMC is significantly affected by the notorious sign problem \cite{Troyer2005}. The sign problem manifests itself as an exponential increase in the number of Monte Carlo (MC) sweeps required to reach convergence \cite{chowdhury2020emulating}. The origin of the problem is that qubit wavefunctions can destructively interfere in the Hilbert space. Quantum problems that do not pose a sign problem are given a special name stoquastic and it is believed non-stoquasticity is an essential ingredient for adiabatic quantum computing (AQC) to be universal \cite{Aharonov2007} and to provide significant speedup over classical computers ~\cite{Vinci2017,Albash2018_2}.
Recently in \cite{King2021}, King et al. demonstrated that with a physical quantum annealing (QA) processor, it is possible to achieve $3$ million times speed up with scaling advantage over an optimized cluster-based continuous time (CT) path integral Monte Carlo (PIMC) code simulated on CPU. In a surprising demonstration (\cite{Isakov2016,Andriyash2017}), King et al. applied the Transverse Field Ising (TFI) Hamiltonian on a geometrically frustrated lattice initialized with a topologically obstructed state. Note that this is a different type of obstruction than the one commonly discussed in the related literature  (see \cite{Hastings2013} for example). This obstruction makes it difficult for an algorithm based on local update schemes to escape the obstruction, whereas a quantum annealer might help escape the obstruction faster. This is interesting because until this result, results on TFI, a well-known stoquastic Hamiltonian, have been routinely benchmarked with quantum Monte Carlo algorithms \cite{denchev2016computational} with no clear scaling differences  for practical problems \cite{albash2018demonstration}. In the theoretical computer science community, the possibility of obtaining a scaling advantage for AQC with sign- problem-free Hamiltonians (such as TFI) is still being actively discussed 
\cite{hastings2021power,gilyen2021sub}. 

PIMC, one of many variations of QMC, is the state-of-the-art tool for simulating and estimating the equilibrium properties of these quantum problems. Powerful and efficient cluster-based algorithms exist for ferromagnetic spin lattices \cite{Rieger1999}. However, it is known that the efficiency of the cluster algorithms drops when frustrations are introduced in the lattice although alternative approaches that compromise between local and global updates were explored \cite{Kandel1990} in the context of the classical Ising model.  

 In recent years, a lot of new devices and domain-specific hardware have emerged to augment the performance of classical computing/simulations in stark contrast with building quantum computers: which is a complete paradigm shift.  In this paper, we explore the possibility of hardware accelerating QMC with one such technology that exploits classical and probabilistic resources, namely, a processor based on probabilistic bits (p-bits) which can be viewed as a classical counterpart of the QA processor \cite{camsari2019scalable}. A p-bit is a robust, classical, and room-temperature entity that continuously fluctuates between two logic states and the rate of this fluctuation can be controlled via an input signal applied to a third terminal \cite{camsari2017stochastic}. p-bits can also be made very compact and can provide true randomness (important for the problem we address in this paper, see Supplementary Note~5) instead of pseudo-random generators, commonly used in software-based solutions. First appeared as a hardware realization of a binary stochastic neuron in \cite{camsari2017stochastic}, later a proof-of-concept p-computer was first demonstrated in \cite{borders2019integer}. p-bit-based hardware solutions have been proposed to improve performance for optimization problems \cite{aadit2022massively}, classical and quantum Monte Carlo \cite{Jan2021Benchmarking}, Bayesian inference \cite{Faria2021Bayesian} and machine learning \cite{Jan2022hardware}. 

\begin{figure*}[!ht]
	\centering
	\vspace{0pt}
	\includegraphics[width=0.8\textwidth,keepaspectratio]{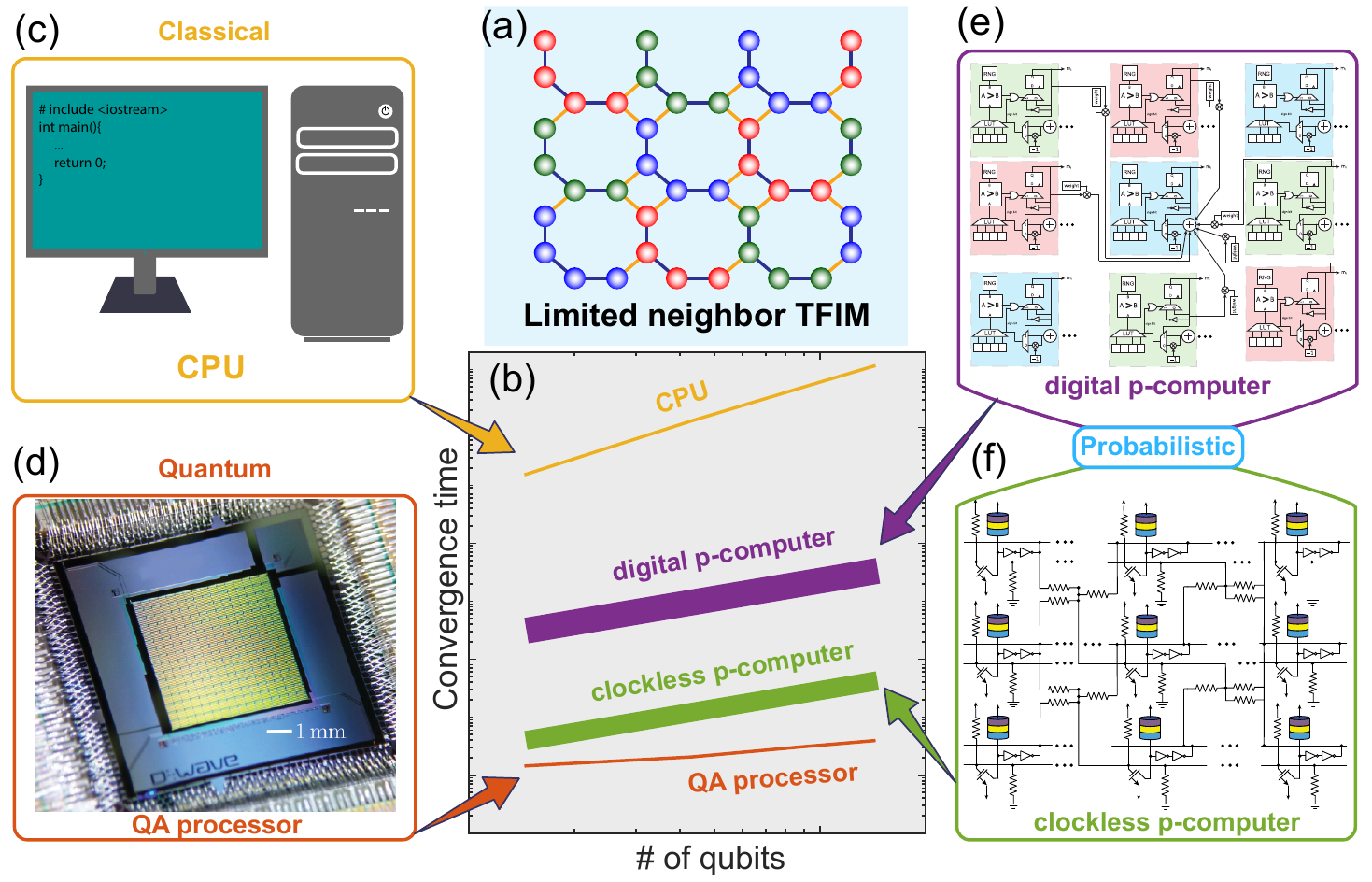}
	\caption{\footnotesize\textbf{Performance of specially designed p-computers, digital and clockless, relative to a CPU and a quantum annealer:} (a) We use an example problem consisting of a lattice of qubits described by a transverse field Ising model. We simulate it classically using the Suzuki-Trotter transformation and calculate a pre-defined order parameter using three different types of hardware whose relative convergence times are sketched in (b). The variations in performances due to variations in implementation technologies are indicated with thick lines. The four types of hardware are also shown schematically -- (c) a von Neumann machine (CPU) -- which simulates the problem by breaking down the problem into a series of instructions and executing them sequentially one after another, (d) a physical quantum annealing processor (QA) that maps the problem onto an interconnected network of rf-SQUIDs (radio-frequency-superconducting quantum interference devices) emulating qubits and rf-couplers coupling those qubits (e) a digital p-computer built using field programmable gate array (FPGA) to lay out a  spatial network of interconnected probabilistic p-bits and (f) a clockless p-computer constructed by interconnecting a network of p-bits through resistors. The quantum annealing processor image has been taken from  Harris, R. et al. Phase transitions in a programmable quantum spin glass simulator. Science 361, 162–165 (2018). Reprinted with permission from AAAS.} 
	\label{fig:hardware_overview}
\end{figure*}

In this study, we demonstrate hardware acceleration using p-bits by an optimized probabilistic computer. This system employs the discrete-time (DT) Path Integral Monte Carlo (PIMC) approach, using the Suzuki-Trotter approximation, and features a sufficient number of replicas to ensure satisfactory accuracy. This design uses massive parallelism and suitable synapses to maximize the number of sweeps collected per clock cycle, resulting in a three-order-of-magnitude improvement in convergence time on a moderately sized programmable gate array (FPGA) compared to a CPU. This design strategy also enables the easy translation of the digital circuit into a clockless mixed-signal design featuring fast resistive synapses and low barrier magnet (LBM) based compact p-bits. Using SPICE (simulation program with integrated circuit emphasis) simulations grounded in experimentally benchmarked models, we anticipate an additional two to three orders of magnitude speedup. Fig.~\ref{fig:hardware_overview} summarizes our approach, illustrating the four different hardware types and their expected relative performances. Overall, our demonstration offers a roadmap for achieving five to six orders of magnitude acceleration for a transverse field Ising model (TFIM) and has the potential to extend to other QMC models as well. The clockless analog hardware can be considered the classical counterpart of the quantum annealer, delivering performance within a factor of less than 10 of the latter. The convergence time for the clockless analog hardware scales with the number of qubits as approximately $N$, which is superior to the $\sim N^{2}$ scaling for CPU implementations but appears worse than that reported for quantum annealers.

\section{Results and Discussion}
\label{sec:results}

\begin{figure}[!ht]
	\centering
	\vspace{0pt}
	\includegraphics[width=0.92\columnwidth,keepaspectratio]{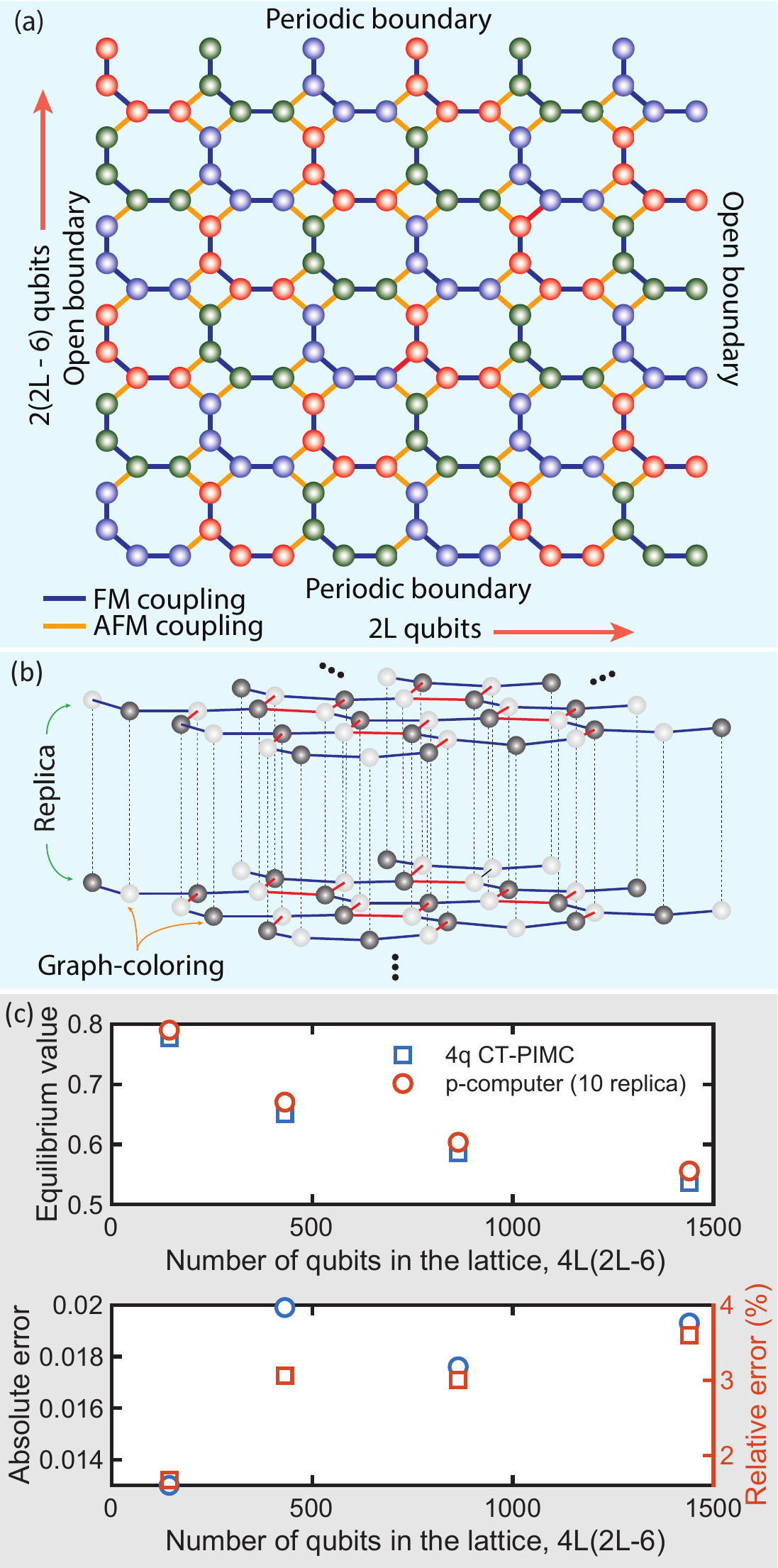}\\
	\caption{\footnotesize\textbf{Example problem addressed in this paper following King et al. \cite{King2021}:} (a) quantum problem solved on quantum annealing processor involves a two-dimensional square octagonal lattice of qubits having $2L$ qubits in one direction and $2(2L-6)$ qubits in the other direction (illustration shows $L = 6$). The blue bond between two qubits denotes ferromagnetic (FM) coupling ($J_{\text{FM}} = -1.8$) and yellow bond indicates antiferromagnetic (AFM) coupling ($J_{\text{AFM}} = 1.0$). The AFM couplings at the open boundary have $J_{\text{AFM}} = 0.5$ (b) Trotterized mapping solved on classical computers, (c) Trotter error with 10 replicas: (Upper panel) Equilibrium values predicted from 10 replica probabilistic computer emulation and four qubit continuous time path intergral Monte Carlo (4q CT-PIMC) algorithm developed in \cite{King2021} (red hollow circles: p-computer data, blue hollow squares: 4q CT-PIMC data). (Lower panel) Absolute (blue hollow circles) and relative errors (red hollow squares) in predicting equilibrium values between the two methods are shown.}
	\label{fig:SQ_OCT_lattice}
\end{figure}

We emulate a quantum problem from a recent work \cite{King2021} where a Transverse Field Ising Hamiltonian (which is stoquastic) 
\begin{equation}
	\mathcal{H_{\rm Q}}=-\left(\sum_{\langle ij \rangle}{J_{ij}\sigma_i^{\rm z}\sigma_j^{\rm z}}+\Gamma\sum_{i}{\sigma_i^{\rm x}}\right)
	\label{eq:TFIM_ham}
\end{equation}
is applied over a two-dimensional square-octagonal qubit lattice as shown in Fig.~\ref{fig:SQ_OCT_lattice}(a). The exotic physics offered by this qubit lattice is of practical interest and has been described in \cite{King2021,King2018}. The square-octagonal lattice can be viewed as a $(2L-6)\times L$ antiferromagnetically (AFM) coupled triangular lattice with a four ferromagnetically (FM) coupled spin basis, giving rise to a total of $4L(2L-6)$ qubits in the lattice. The resulting lattice consists of square and octagonal plaquettes which are periodically connected along one direction and it has open boundaries in the other direction. In the bulk of the lattice, each qubit is connected to three other neighbors whereas, at the open boundary, some qubits are connected to just one neighbor, and others are connected to two neighbors.  To increase the degeneracy of the classical ground state, the AFM couplings at the open boundary are also reduced to half of that in the bulk.

Each square or octagonal plaquette in this lattice is composed of qubits from three different sublattices and has three (an odd number) AFM bonds (for both octagonal and square plaquettes). This leads to a frustrated lattice since it is impossible to satisfy all the bonds simultaneously. Three different qubit sublattices within the lattice are indicated by the red, green, and blue colors in Fig.~\ref{fig:SQ_OCT_lattice}(a).

In this benchmark study, we observe the average equilibration speed of the average order parameter when initialized with a particular classical state (in this study we will be referring to two particular initial states: counterclockwise (CCW) and ordered, see Supplementary Note~1 for more details) in probabilistic computer which is based on discrete-time path integral Monte Carlo (DT-PIMC) with many interconected replicas of the original qubit lattice but the qubits are replaced by p-bits (see Methods section). We will compare this result against the general-purpose processor (CPU) and with the quantum annealing processor from \cite{King2021}. The procedure to obtain the average order parameter was defined in \cite{King2021} and has been outlined in the Methods section.

\subsection{Design considerations for the probabilistic emulator}
\label{sec:pcomp_design}

We start the process of designing our p-computer with the trotterization of the qubit lattice using 10 replicas and involving $40L(2L-6)$ p-bits, ranging up to 14,400 p-bits for $L=15$.  Traditional Gibbs sampling or single-flip Monte Carlo sampling takes too long to converge for such a large network and we need a scheme that allows us to simultaneously update many p-bits. But it is also well-known that updating two p-bits simultaneously which are connected to each other, leads to erroneous output. We realized that the limited connections among the p-bits in the replicated network could be utilized to achieve massive parallelism where many p-bits can be updated in parallel and therefore can be used to speedup the convergence. To obtain such massive parallelism, we next applied graph coloring on the replicated p-bit network, as recently explored in  Ref.~\cite{aadit2022massively} for general and irregular lattices.

Graph coloring assigns different colors to p-bits that are connected to each other and ensures that no two p-bits that are connected to each other have the same color thus enabling us to update all p-bits in the same color group simultaneously. Although not immediately obvious to many, it can be easily checked that the qubits of the square-octagonal lattice under consideration can be colored using just two colors (i.e., the lattice is bipartite). If we always choose an even number of replicas (which is what we do in this work), then we found that the translated p-network can also be colored using just two colors (i.e., the p-bit network also remains bipartite) as shown in Fig.~\ref{fig:SQ_OCT_lattice}(b). Hence with just two colors, half of the p-bit network can be updated in one clock cycle and the other half of the network in another, producing one sample in every two clock cycles. In general, compared to a single flip Monte Carlo implementation which updates one spin in one clock cycle, this graph-colored approach can reduce the number of clock periods required to converge by a factor of $\sim nr/C$ ($nr$ is the number of p-bits and $C$ is the number of colors) assuming same clock period for both cases.

This leads us to argue that a p-computer should exhibit weaker dependence with the increasing size of the network compared to the CPU because even though the number of p-bit increases in the network, one can also proportionally increase the number of p-bits (we estimate that up to one million of p-bits can be integrated on a chip with a reasonable power budget \cite{sutton2020autonomous}) to be updated in a given clock, yielding a factor of $n$ ($=$ number of p-bits in the network) improvement in scaling over CPU. This demonstrates the power of a properly architected p-computer over a CPU where the scope of such parallelization is very limited.

\subsection{Results from digital p-computer emulation on FPGA}

To demonstrate the utility of such massively parallel architecture, we next emulate this graph-colored p-bit network by implementing it on FPGA using Amazon Web Services F1 instance (more details of the FPGA implementation can be found in Supplementary Note~2).  Various implementations of p-bits including digital and analog have been discussed in \cite{sutton2020autonomous,schowdhuryIEDM2019}. The digital implementations of p-bits are costly in terms of resources and require thousands of transistors per p-bit and so we have only been able to fit the smallest lattice size ($L=6$) with the resources provided therein. But we expect that when replaced with nanomagnet-based stochastic MTJs, the situation would improve drastically. It is also equally important to carefully design the synapse that can provide updated information to p-bits by quickly responding to any changes in the state of neighboring p-bits. In the spirit of \cite{Kaiser2021}, we carefully choose our synapse to update $nrf/C$ p-bits per second providing $f/C$ sweeps per second. The clock period $1/f$ needs to be minimized carefully so that the synapse can correctly calculate the response while providing maximum throughput. This choice of FPGA implementation also provides a unique way that permits a clear pathway to a mixed signal circuit. The FPGA is less than an ASIC (application-specific integrated circuit), but the mixed signal especially with MTJs would be much more than an ASIC. 

In our FPGA demonstration, we have been able to run the smallest lattice with an 8 ns clock period (16 ns per sweep since we have two colors) and we believe that given enough resources we should also be able to run the bigger lattices at the same clock frequency. We project convergence times for other lattice sizes based on CPU simulations. These `projections' are based on actual implementation with real devices and should be reliable, given our digital architecture and the fact that we did not use the largest FPGA available today.

\begin{figure}[!ht]
	\centering
	\vspace{10pt}
 \includegraphics[width=0.9\columnwidth,keepaspectratio]{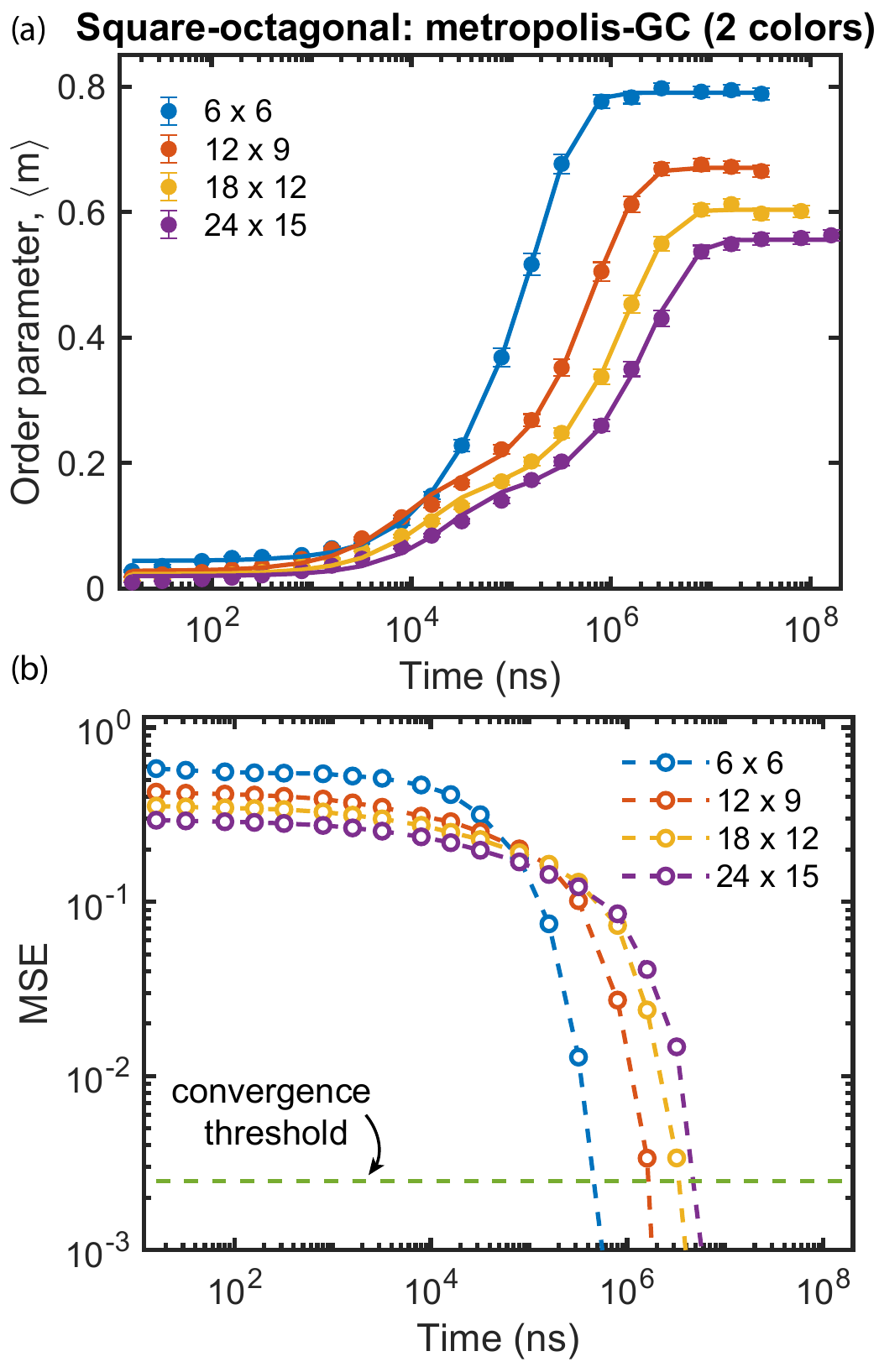}
	\caption{\footnotesize\textbf{Extracting convergence times from simulation results following the protocol described in \cite{King2021}:} (a) Average order parameter $\langle m\rangle$ as defined in Eq.~(\ref{eq:orderParam}) (Eq.~(2) in \cite{King2021}) are obtained for four different lattice sizes (blue: $6\times6$, red: $12\times9$, yellow: $18\times12$, purple: $24\times15$) using the mapped p-bit network. We have used $\Gamma = 0.736$, and $\beta = 1/0.244$ for which the scaling difference was reported to be maximum.  We show results only for the counterclockwise wound initial condition as explained in the Supplementary Note~1. Only the curve labeled as $6\times 6$ is obtained through actual field programmable gate array emulation and the others are projected based on CPU simulation. All data points are averaged over 1000 different runs and the errorbars correspond to $95\%$ confidence interval around the mean. Filled circles represent data points while the solid lines represent $a\exp{(-bx)}+c\exp{(-dx)+g}$ type fit. (b) Mean squared error (MSE) plot for each lattice size (blue: $6\times6$, red: $12\times9$, yellow: $18\times12$, purple: $24\times15$; hollow circles: data, dashed lines: fit) calculated from their corresponding `$g$' values in (a). The scaling is more clearly visible in this plot. Also shown is the 0.0025 threshold in dashed green which is used to define convergence.}
	\label{fig:AQCResults}
\end{figure}

For the other lattice sizes, we obtain the average order parameter versus the number of sweeps plots via running MATLAB on CPU (a verification of FPGA output matching MATLAB output is also provided in Supplementary Note~2) and then multiply the $x$ axis of that plot by 16 ns per sample. These lead to the curves in Fig.~\ref{fig:AQCResults}(a), where we report the average order parameter, $\langle m(t)\rangle$ versus time curves obtained from p-computer emulation for the same four lattice sizes of square-octagonal lattice and with the same parameters as in \cite{King2021} but only with counterclockwise (CCW) wound initial condition. The curves with clockwise wound (CW) initial condition are similar to these CCW curves (slightly faster than CCW) whereas curves for ordered initial condition (not shown) show much faster convergence. 

In Fig.~\ref{fig:SQ_OCT_lattice}(c), we report the error in predicting the saturation value from using finite replica in our p-computer emulations. We compare our results against the 4q CT-PIMC algorithm developed in \cite{King2021}. To ensure fidelity, we use the same C++ codes provided therein. With 10 replicas, we reproduce  the CT-PIMC results with an  absolute difference of $0.01\sim0.03$ from the smallest to the largest lattice sizes (see Supplementary Note~4 for CT-PIMC results). As reported in  Ref.~\cite{King2021}, we do not observe systematic changes in Trotter errors with lattice sizes.

\begin{figure*}[!ht]
	\centering
	\vspace{0pt}	\includegraphics[width=0.9\textwidth,keepaspectratio]{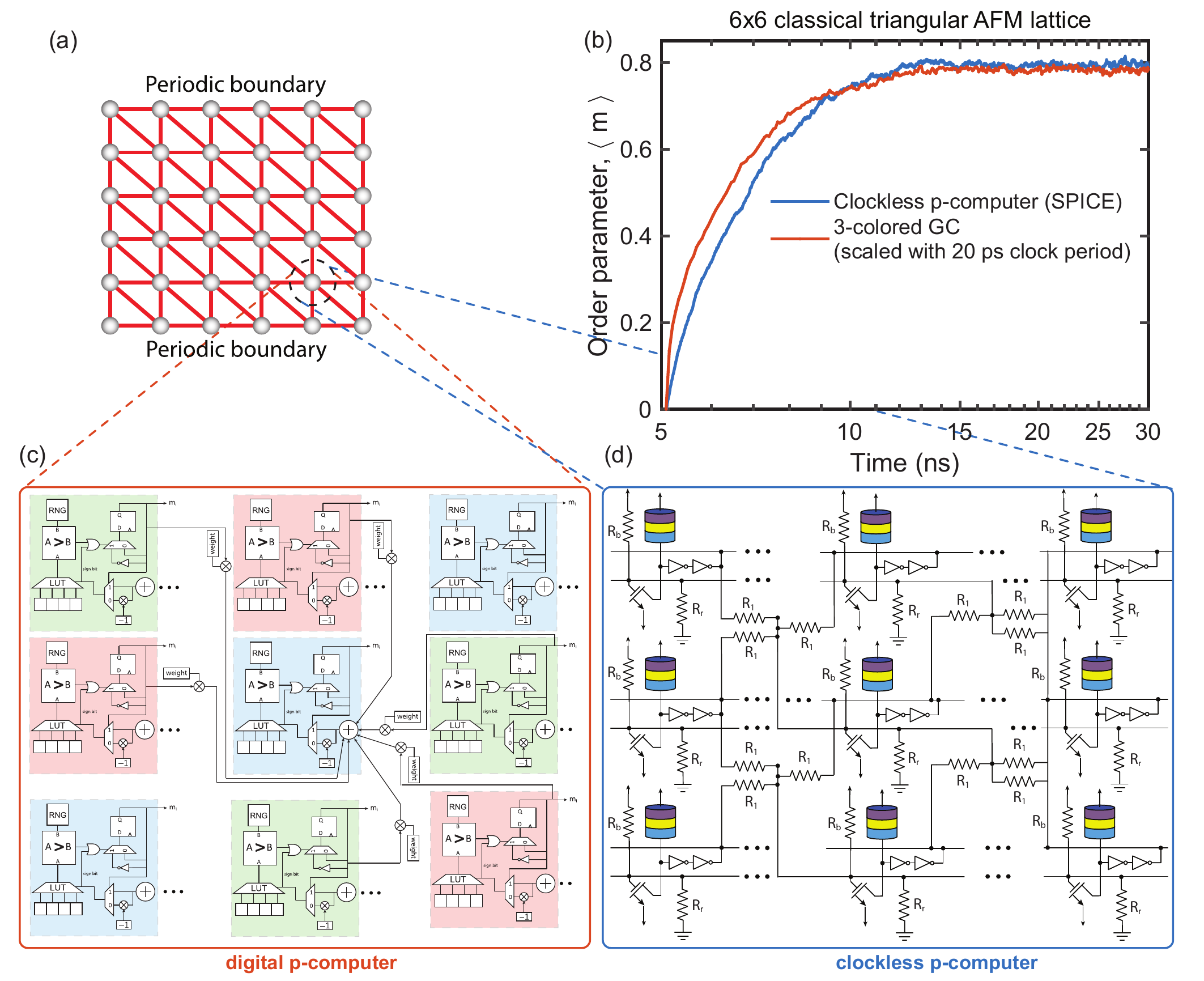}
	\caption{\footnotesize\textbf{Comparing digital to clockless p-computer}: (a) A $6\times6$ antiferromagnetically (AFM) coupled triangular lattice with classical spins is shown. (b) The convergences of the order parameter for the lattice shown in (a) are plotted for two different p-computer design approach (red solid line: graph-colored based digital design and  blue solid line: nanomagnets based analog design) discussed in this work. We have used $\beta=2$ in this example. (c) The graph-colored (GC) based digital p-computer design where the convergence is estimated from MATLAB simulations assuming $c \times f_{c}$ ($c=3$ for triangular lattice) sweeps are collected every per second, $f_{c}$ being the clock frequency. (d) The clockless p-computer design which is simulated using SPICE simulator.}
	\label{fig:TRI_AFM}
	\end{figure*}

\subsection{Clockless Autonomous Operation}
\label{sec:autonomous}

In the last subsection, we have presented a digital implementation of a p-computer based on the graph-colored architecture. Even though it is not immediately obvious to many, in that architecture, we managed to use just two colors which happen to be the minimum number of colors possible and thus maximizes the number of p-bits that can be updated simultaneously. This allowed us to greatly reduce the convergence time compared to single-flip Monte Carlo which updates just one p-bit at a given clock period and thus converges very slowly. However, there are two problems associated with this graph-colored digital implementation: first, a fully digital implementation of a p-bit requires thousands of transistors which increases the hardware footprint per p-bit quite significantly. This can be mitigated somewhat through the use of nano-magnet-based compact p-bits which uses just three transistors and an MTJ. However, this also requires the use of digital to analog converters for each p-bit since the input to such compact p-bits is analog. The second issue with the digital implementation is that to perform a colored update, all p-bits need to be synchronized through a global clock, the distribution of such clock throughout the chip becomes complicated with the increasing number of p-bits and also slows down the frequency with which the system can be operated.

To circumvent the above issues, we next visit a fully analog implementation of a p-computer with a clockless autonomous architecture. The clockless architecture is inspired by nature: natural processes do not use clocks. In clockless autonomous architecture, we do not put any restrictions on the updating of p-bits. Each p-bit can attempt to update at any point in time without ever requiring a clock to guide them. Of course, errors will be incurred if two connected p-bits update themselves simultaneously, and therefore with this scheme, it is essential to minimize the probability of happening that. If there are $d$ neighbors to each p-bit then the probability that two connected p-bit will update simultaneously is roughly $d\times s^2$ where $s=\tau_s/\tau_N$, $1/\tau_N$ is the frequency with which a p-bit attempt to update itself and $\tau_s$ is the time required to propagate the information of a p-bit update to its neighbors. To make this clockless autonomous operation work it is essential to have $s\ll1$ (usually $s\approx 0.1$ works well). This interesting possibility of clockless autonomous operation was introduced in \cite{sutton2020autonomous} where a digital demonstration was made using FPGA. However, in this work, we use a simple resistive synapse-based architecture. Since resistors can instantaneously respond to the change in applied voltage, this type of synapse should be very fast compared to the average fluctuation time of s-MTJ-based p-bits ($\sim 100$ ps). We demonstrate the validity of this scheme by showing a SPICE simulation of a $6\times6$ triangular AFM lattice with classical spins as shown in Fig.~\ref{fig:TRI_AFM}(a). As mentioned earlier, the triangular lattice is the base lattice of the square-octagonal lattice we have used so far.  A partial view of the analog circuit simulated in SPICE which corresponds to the lattice above is also shown in Fig.~\ref{fig:TRI_AFM}(d). We only show the resistive analog synapse providing the input for a single p-bit as marked. We use similar parameter values and the same boundary conditions as we have used for the square-octagonal lattice (the same AFM coupling strength (\(\vert J_{AFM}\vert=1\)) inside the lattice and \(\vert J_{AFM}\vert=0.5\) at the open boundaries). We also use the same definition for the order parameter. To keep it similar to what we have done in the previous section, we also use CCW initial condition in this example. Doing these help us to solve the problem in SPICE within a reasonable amount of time. 

Fig.~\ref{fig:TRI_AFM}(b) shows the relaxation of the order parameter with time for the example described in Fig.~\ref{fig:TRI_AFM}(a). We use the same SPICE p-bit model used in \cite{Camsari_EDL}. We also show the relaxation curve obtained via a 3-graph-colored architecture (the triangular lattice in this example is 3-colorable). The graph-colored system as shown in Fig.~\ref{fig:TRI_AFM}(c) converges (based on the criterion we have used so far) around 72 sweeps. In 125 MHz FPGA that we have used earlier, this would take around \(72\times3\times8\) ns \(= 1.73\) $\mu$s, whereas the corresponding analog circuit implementation converges in around 5 ns, converging around 400 times faster than similar digital implementation used earlier. Although the circuit used here is not programmable it nicely illustrates the principle that around two orders of additional speed-up can be obtained with the use of a properly designed fully analog and clockless p-computer. 

\begin{figure}[!t]
\centering
\vspace{0pt}
\includegraphics[width=0.9\columnwidth,keepaspectratio]{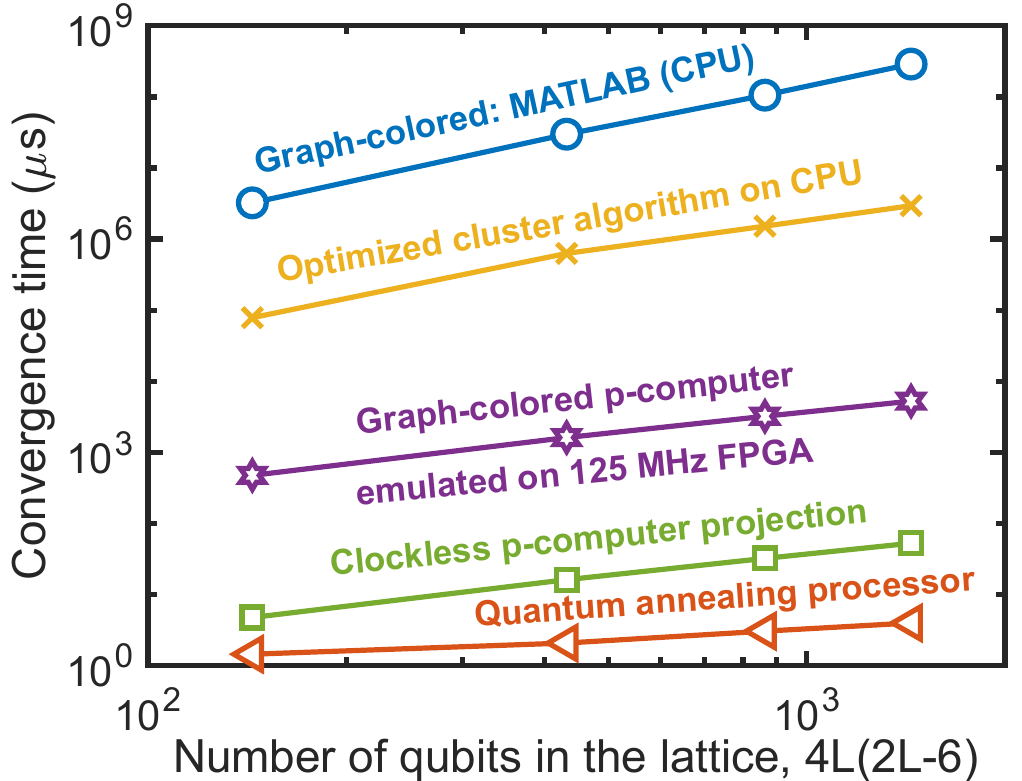}
\caption{\footnotesize\textbf{Convergence time for problems of increasing size using different hardware implementations:} The convergence time is extracted using the protocol described in Data fitting and convergence criterion subsection of Method section for the problem defined in Fig.~\ref{fig:SQ_OCT_lattice}. The data for optimized CPU (yellow cross) (mean of counterclockwise and clockwise wound initial states) and quantum annealing processor (red triangles) (for counter clockwise wound initial state only) are extracted from the information provided in King et al.  \cite{King2021} (Fig.~4 and supplementary Fig.~13 therein), while the results for graph-colored MATLAB (blue circles) (simulated on CPU), digital (purple star) and clockless p-computers (green squares) are obtained in this paper as described in the text. This CPU simulation of graph colored algorithm is performed with a vectorized MATLAB code running on Intel(R) Core(TM) i7-10750H CPU (2.60 GHz, 16 GB RAM, Windows 11 Home operating system and single thread simulation are used). As mentioned in the caption of Fig.~\ref{fig:AQCResults}, only the leftmost data point on the p-computer (FPGA: field programmable gate array) curve is obtained from actual FPGA emulation. The rest of the points on the curve are projections based on the simulations on the CPU. The slope of the p-computer curve is clearly different (and smaller) from the graph-colored MATLAB (CPU) and optimized four qubit continuous time path integral Monte Carlo algorithm (simulated on CPU) curves (the exact slopes found using curve fitting are listed in Supplementary Note~7). }
	\label{fig:timeScaling}
\end{figure}

\subsection{Convergence time scaling results} Finally, we show the time scaling for four different hardware in Fig.~\ref{fig:timeScaling}. We directly adopt the optimized 4q CTPIMC (CPU) and QA processor data from \cite{King2021}. A simple curve fitting to CPU data reveals a roughly $N_Q^2$ scaling where $N_Q = 4L(2L-6)$ is the total number of qubits in the lattice. On the other hand, the p-computer results show a prefactor improvement and an improvement in scaling compared to a CPU. For a more direct comparison, we also show the scaling of our graph-colored algorithm simulated on CPU which also shows an $\sim N_Q^2$ scaling behavior. We observe an $\sim N_Q$ scaling for p-computer and as noted before, the reason for such a scaling improvement is due to the exploitation of massive parallelism where the number of p-bits that can be updated also increases with the lattice size and this is not due to an algorithmic improvement (the scaling with the number of p-bits is provided in Supplementary Note~6). 

In our perspective, any CPU-based solution is unlikely to achieve the same level of parallelism as our p-computer, as this would necessitate the use of ``$N$'' processors or threads. While we acknowledge that specialized CPU implementations employing multiple threads and/or processors might approximate the parallelism achieved with our custom hardware, our optimized implementation suggests that achieving such parallelism (scaling with N-threads) is not trivial. Additionally, beyond digital implementations, nanodevice-based ASICs could support millions of p-bits \cite{sutton2020autonomous}, taking $N$ to unprecedented levels. This degree of parallelism may be challenging to replicate in conventional digital hardware, at least from a practical standpoint.

We note here that we are investigating a quantum \textit{sampling} problem in this work that measures the equilibration time of a specially prepared lattice. The measured convergence time does not depend on the number of replicas run in parallel. This is because the reported time is the `average' obtained from $R$ identically prepared trials. Whether these trials are taken in parallel or in series does not affect the convergence time we (or King et al.) report. Increasing the number of replicas simply reduces the variance of the random variable we are estimating and has no effect on the reported wall-clock times or convergence times in any of the platforms (annealer, CPU, p-computer).  For the same reason, we do not fit as many parallel replicas in our FPGA as possible to make our measurements, even for  smaller lattice sizes.

On the contrary, in optimization-type problems where the quantity of interest is ``time to solution'' (TTS), it makes sense to utilize a p-bit/qubit system to the fullest by running as many parallel instances as possible in one run (to reduce TTS linearly by reducing the number of repetitions necessary). In such cases, appropriate care must be taken when comparing the performance of specialized hardware with CPU performance which is usually utilized fully (see \cite{Roennow2014} for example). 

Our results for the digital implementation of p-computer emulated on 125 MHz FPGA show that for the largest lattice size 
($L=15$) that has been emulated in \cite{King2021}, we should get a $\sim 1000\times$ improvement over a single thread implementation on CPU. But it stands, the current FPGA emulations of our p-computer are \(\sim 3\) orders of magnitude worse than the physical quantum annealing processor. We expect another one order of magnitude improvement might be possible with this approach by using a customized mixed-signal ASIC design with stochastic magnetic tunnel junction (sMTJ)-based p-bits. 
However, based on the example of clockless operation shown in Section~\ref{sec:autonomous},  we project another two orders of magnitude improvement in convergence time. This brings the gap with the quantum annealing processor down to one order or less. The operation of the quantum annealing processor might be governed by non-local quantum processes leading to the $\sqrt{N_Q}$ scaling predicted in \cite{Montanaro2015}, though there are not enough data points to be certain.

Although we did not do a direct GPU (graphics processing unit) implementation of the problem under consideration, we looked for the GPU emulations of bipartite (2-colorable) graphs (like the one being simulated in this work) in the literature. In a typical GPU one gets around 10-30 flips/ns (the key metric used to compare the performance and the higher flips/ns gives better performance) \cite{fang2014parallel,yang2019high,BLOCK20101549,Preis2009GPU} for such graphs whereas our designs get 90 flips/ns (1440/2 = 720 p-bits being flipped at every 8 ns) from the actual FPGA design for the smallest lattice size and will increase as we enable ourselves to integrate more and more p-bits.

\section{Conclusion}

In this work, we have presented a roadmap for hardware acceleration of QMC which is ubiquitously used in the scientific community to study the properties of many-body quantum systems. We have mapped a recently studied quantum problem into a carefully designed autonomous probabilistic computer and projected 5-6 orders of magnitude improvement in convergence time which is within a factor of $10$ of what has been obtained from a physical quantum annealer. The massively parallel operation of a probabilistic computer together with the clockless asynchronous dynamics provides a significant scaling advantage compared to a CPU implementation. Robustness, room-temperature operation, low power consumption, and ultra-fast sampling -- these features make it interesting to investigate the applicability of probabilistic computers to other quantum problems beyond the TFI Hamiltonian studied in this work.

\section{Methods}
\subsection{Procedure to calculate average order parameter}
For the sake of completeness, we provide the details of the calculation of average order parameter in the following:
\begin{enumerate}
	\item Average of four FM-coupled qubits is computed for each basis in the lattice. Depending on the sublattice the basis belongs to, these averages are denoted as $m_{\text{av,red}}$, $m_{\text{av,green}}$ or $m_{\text{av,blue}}$ (see Fig.~\ref{fig:SQ_OCT_lattice}). As mentioned earlier, averaging over basis turns the lattice into an AFM-coupled triangular lattice.
	\item For each triangular plaquette in the transformed triangular lattice (including those formed from the periodic boundary), compute the complex-valued quantity known as pseudospin which is defined as follows:
	\begin{equation}
		\zeta_{\rm pl}=\frac{1}{\sqrt{3}}({m}_{{\rm av,red}}+{\rm e}^{2 \pi {\rm   i}/3}{m}_{\rm {av,green}}+{\rm e}^{4 \pi{\rm i}/3}{m}_{{\rm av,blue}}).
		\label{eq:orderParam}
	\end{equation}
	\item Average over all triangular plaquettes, i.e.,
	\begin{equation}
		\zeta_{\rm conf} = \frac{1}{N_{\text{pl}}}\sum_{i}{\zeta_{{\rm pl},i}},
		\label{eq:pseudospin}
	\end{equation}
    where $N_{\text{pl}}$ is the number of plaquettes (including periodic boundaries in the quantum lattice).
	\item Obtain the average order parameter by taking the average of absolute values for different configurations of the lattice, i.e.,
	\begin{equation}
		\langle m\rangle = \sum_{k}{p_k\left\vert\zeta_{{\rm conf},k}\right\vert},
        \label{eq:avgOrderParam}
	\end{equation}
	where $p_k$ is the probability of occurrence for configuration $k$.
\end{enumerate}

\subsection{Discrete-time Path Integral Monte Carlo}
\label{sec:pcomp_algo}

Our p-computer is a discrete-time path integral Monte Carlo (DT-PIMC) emulator based on the Suzuki-Trotter approximation \cite{suzuki1976relationship}. The idea of such a hardware emulator for QMC was first proposed in \cite{camsari2019scalable}. In this scheme, one tries to approximate the partition function of the quantum Hamiltonian, $Z_{\rm Q}$:
\begin{equation}
	Z_{\rm Q} = \text{tr}\left[\exp{\left(-\beta H_{\rm Q}\right)}\right]
\end{equation}

\noindent with a classical Hamiltonian, $H_{\rm Cl}$ such that the partition function corresponding to $H_{\rm Cl}$ is equal to $Z_{\rm Q}$. For the quantum Hamiltonian in Eq.~(\ref{eq:TFIM_ham}), one finds that the following classical Hamiltonian, $H_{\rm Cl}$:
\begin{equation}
H_{\rm Cl}=-\sum_{k=1}^{r}{\left[\sum_{i<j}{J_{\parallel,ij}m_{i,k}m_{j,k}}+\sum_{i}{J_{\perp}m_{i,k}m_{i,k+1}}\right]}
\end{equation}
with 
\begin{eqnarray}
J_{\parallel,ij}&=&J_{ij}/r,\\
J_{\perp}&=&-(0.5/\beta)\ln{\left[\tanh{\left(\beta\Gamma/r\right)}\right]}
\end{eqnarray}
and $m_{i,j}\in\{-1,+1\}$ yields the same
$Z_{\rm Q}$ in the limit $r\to\infty$. The error goes down as $\mathcal{O}(1/r^2)$ and in practice, one can find a reasonably good approximation with a finite number of replicas in many cases. 

\subsection{Data fitting and convergence criterion} Each curve in Fig.~\ref{fig:AQCResults}(a) is then fitted with $ae^{-bx}+ce^{-dx}+g$ type fitting model (a justification for using this fitting model is provided in the Supplementary Note~3) where $g$ represents the prediction for equilibrium value of average order parameter from p-computer emulation. Fig.~\ref{fig:AQCResults}(b) shows the decay in mean squared error (MSE) as time increases. It also clearly shows that the time required to reach a fixed MSE level increases as the size of the lattice increases. We define convergence time as the time required to reach an MSE level of $0.0025$, which is equivalent to finding the time required to reach $g-0.05$ in Fig.~\ref{fig:AQCResults}(a) and was used to define convergence in \cite{King2021}.

\subsection{Averaging over sweeps from parallel runs to avoid autocorrelation} We note that to get the true average convergence time of the p-bit network, we run each lattice emulation many times each time with different seed in random number generator and compute the average order parameter at each time point by taking an average of the absolute value of the order parameter calculated at the same time point from all the runs only. This allows us to eliminate the correlation between sweeps taken from the same run which yields longer convergence times and does not represent the actual convergence time of the network. 

\subsection{More about the implementation of the clockless p-computer circuit} To simulate the analog circuit in Fig.~\ref{fig:TRI_AFM}(d), we have used a simple voltage divider based synapse with $R_0 = 15 \text{M}\Omega$, $R_r = R_0/10$, $R_b = R_0/4$ (for bias inputs) and $R_1 = R_0/3.5$ (for the AFM weight of magnitude 1). For p-bits on the border along horizontal direction (open boundary condition), we have used $R'_1 = R_0/2$ (to represent the AFM weight of magnitude 1) and $R''_1 = R_0$ (to represent the AFM weight of magnitude 0.5).

\section{Data Availability Statement}
The data used for generating the figures are available upon request to the author (email: datta@purdue.edu, schowdhury.eee@gmail.com).

\section{Code Availability Statement}
The codes used for generating the figures are available upon request to the author (email: datta@purdue.edu, schowdhury.eee@gmail.com).

\section*{Acknowledgments}
This work was supported in part by ASCENT, one of six centers in JUMP, a Semiconductor Research Corporation (SRC) program sponsored by DARPA. KYC acknowledges support from the Office of Naval Research YIP program. The authors thank Dr. Jan Kaiser and Rishi Kumar Jaiswal for many helpful discussions, especially those related to the optimization of the FPGA implementation. The authors are also grateful to Dr. Brian M. Sutton whose work \cite{sutton2020autonomous} has been used extensively in our p-computer design. S.~C. was with Elmore Family School of Electrical and Computer Engineering, Purdue University, IN 47907, USA when this work was done.

\section{Author Contribution}

S.~C. performed the simulations with help from K.~Y.~C. and wrote the first draft of the manuscript. S.~C., K.~Y.~C and S.~D. contributed to and participated in designing the experiments, analyzing the results and editing the manuscript.

\section{Competing Interests}
S.~D. has a financial interest in Ludwig Computing. The authors declare no other competing interests.

\makeatletter 
\renewcommand{\fnum@figure}{\textbf{Supplementary Figure~\thefigure}}
\makeatother

\makeatletter 
\renewcommand{\fnum@table}{\textbf{Supplementary Table~\thetable}}
\makeatother

\setcounter{section}{0}
\renewcommand{\theequation}{S.\arabic{equation}}

\renewcommand\thesection{\arabic{section}}

\titleformat{\section}{\filcenter\normalfont\small \bfseries}{Supplementary Note \thesection:}{1em}{\MakeUppercase} 

\makeatletter
\def\bbordermatrix#1{\begingroup \m@th
	\@tempdima 4.75\p@
	\setbox\z@\vbox{%
		\def\cr{\crcr\noalign{\kern2\p@\global\let\cr\endline}}%
		\ialign{$##$\hfil\kern2\p@\kern\@tempdima&\thinspace\hfil$##$\hfil
			&&\quad\hfil$##$\hfil\crcr
			\omit\strut\hfil\crcr\noalign{\kern-\baselineskip}%
			#1\crcr\omit\strut\cr}}%
	\setbox\tw@\vbox{\unvcopy\z@\global\setbox\@ne\lastbox}%
	\setbox\tw@\hbox{\unhbox\@ne\unskip\global\setbox\@ne\lastbox}%
	\setbox\tw@\hbox{$\kern\wd\@ne\kern-\@tempdima\left[\kern-\wd\@ne
		\global\setbox\@ne\vbox{\box\@ne\kern2\p@}%
		\vcenter{\kern-\ht\@ne\unvbox\z@\kern-\baselineskip}\,\right]$}%
	\null\;\vbox{\kern\ht\@ne\box\tw@}\endgroup}
\makeatother

\makeatletter
\def\BState{\State\hskip-\ALG@thistlm}
\makeatother

\setcounter{secnumdepth}{1}
\setlength{\belowcaptionskip}{-11pt}



\onecolumn

\begin{center}
	Supplementary Information For:\\
	\vspace{5pt} {\Large \bf Accelerated Quantum Monte Carlo with Probabilistic Computers}\\
	\vspace{5pt}
	Shuvro Chowdhury, Kerem Y. Camsari and Supriyo Datta
\end{center}

\section{Ordered and Counterclockwise wound initial states}
\label{app:chapDWAVE_CCW}

Eq.~(3) in the main article prescribed how to compute average pseudospin given a state of the qubits. Interestingly, we can carefully assign values to our qubits to get a configuration where the value of the average pseudospin is the maximum. This occurs when all the triangular plaquettes inside this configuration align themselves in a particular direction as shown in Supplementary Figure  \ref{fig:ordered_pseudospin} which gives it the name ``ordered" state. The pseudospin value for this configuration is the maximum value for any square or octagonal plaquette i.e., $2/\sqrt{3}$.  We also note that this configuration is a ground state of a square-octagonal lattice of classical spins.
\begin{figure}[!ht]
	\centering
	\includegraphics[width=3in,keepaspectratio]{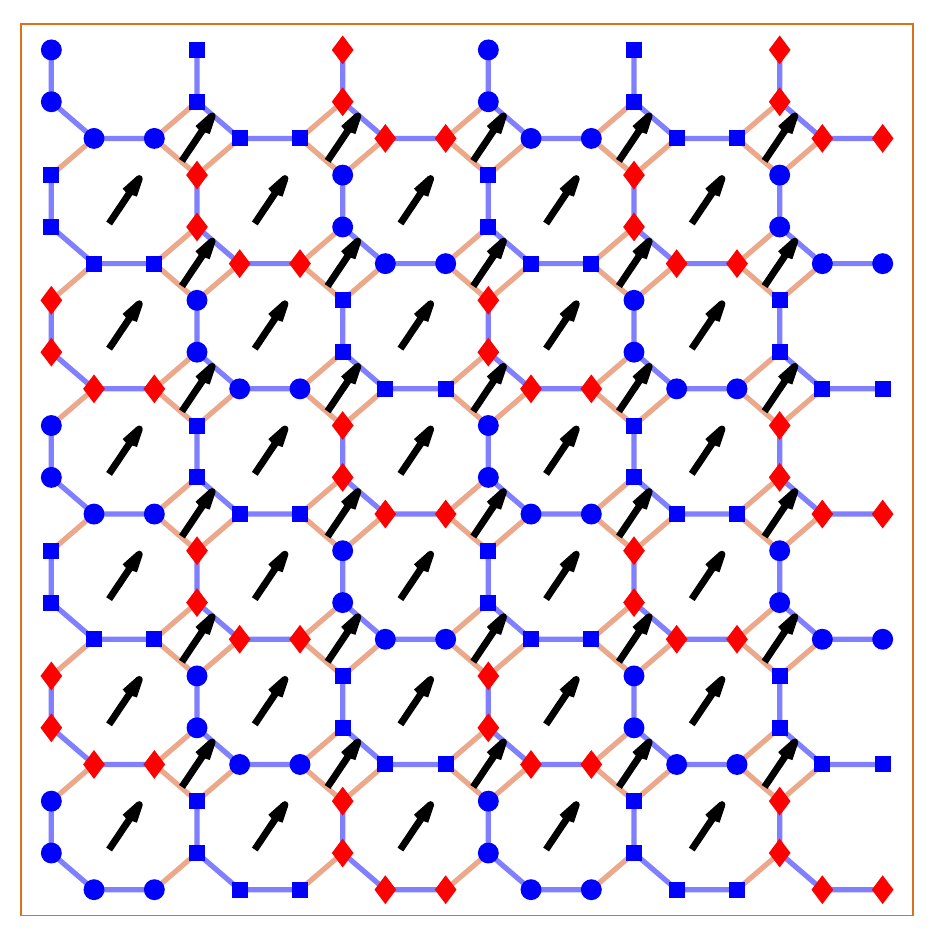}\\
	\caption{A $6\times6$ (144 qubits) 2D square-octagonal lattice in ordered state. Circles denote qubits in the first sublattice, squares denote qubits in the second sublattice while diamonds represents qubits in the third sublattice. Blue color represents qubits in $|1\rangle$ state whereas red color represents qubits in $|0\rangle$ state.  All the plaquettes align themselves at an angle $60^{\circ}$ with the horizontal axis (black arrow). Note that there are six possible directions the plaquettes can align themselves to.}
	\label{fig:ordered_pseudospin}
\end{figure} 

\vspace{0.25cm}

It is not the case that a ground state of the classical square-octagonal lattice always yields the maximum value for pseudospin. It is also possible to have classical ground states for which the pseudospin value is zero (minimum). Two examples of such states are counterclockwise and clockwise wound states as shown  in Supplementary Figure  \ref{fig:CCW_CW}. The pseudospins of individual plaquettes for these configurations do not align themselves but rotate along the periodic boundary direction. 
\begin{figure*}[!ht]
	\centering
	\includegraphics[keepaspectratio,width=0.85\columnwidth]{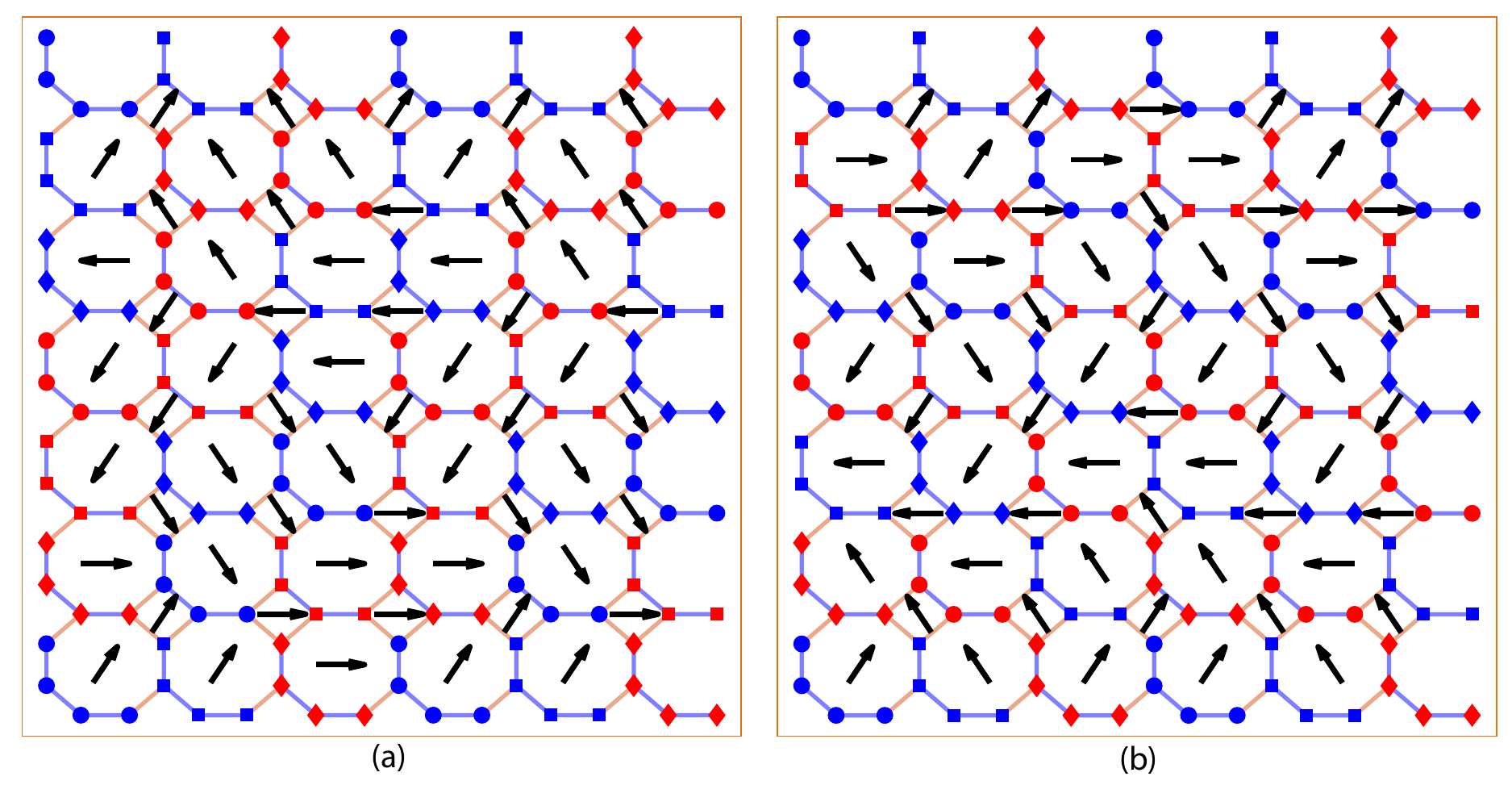}\\
	\caption{A $6\times6$ (144 qubits) 2D square-octagonal lattice in (a) counterclockwise and (b) clockwise wound states. The pseudospin of individual plaquettes rotates along the periodic boundary (vertical) direction. These two states lead to a minimum value (zero) for average pseudospin. Circles denote qubits in the first sublattice, squares denote qubits in the second sublattice while diamonds represents qubits in the third sublattice. Blue color represents qubits in $|1\rangle$ state whereas red color represents qubits in $|0\rangle$ state.}
	\label{fig:CCW_CW}
\end{figure*} 

An interesting observation is that it is not possible to go to the ordered state from these counterclockwise and clockwise wound states just by making a local change in the configuration. The other way of saying this is that these configurations are topologically protected. One needs to change all the spins simultaneously which also makes it difficult for algorithms that create sweeps by proposing local changes in the configuration of the previous sample to quickly converge to the saturation value if the previous sample happens to be in one of these states.

\section{More details of FPGA implementation}
\label{app:chapDWAVE_FPGA_details}

In this work, we define the operation of a p-bit using the following two equations:
\begin{eqnarray}
	\Delta E_i &=& m_i\sum_j{W_{ij}m_j} \label{eq:pbit_synapse}\\ 
	m_i &=& -m_i\,\text{sgn}(e^{-2\beta\Delta E_i}-r_{[0,1]}) \label{eq:pbit_neuron}
\end{eqnarray}
where $j$ in Eq.~(\ref{eq:pbit_synapse}) runs through the set of the neighbors of $i$\textsuperscript{th} p-bit and $r_{[0,1]}$ is a random real number uniformly distributed in $[0,1]$. Eq.~(\ref{eq:pbit_synapse}) is known as the synapse equation and defines the input to the $i$\textsuperscript{th} p-bit. On the other hand, Eq.~(\ref{eq:pbit_neuron}) is known as the neuron equation and defines how a p-bit should change its state given the input. There is another well-known form of synapse-neuron equations in the literature which is:
\begin{eqnarray}
	I_i &=& \sum_j{W_{ij}m_j} \label{eq:pbit_synapse2}\\ 
	m_i &=&\text{sgn}(\tanh{(\beta I_i)}-2r_{[0,1]}+1) \label{eq:pbit_neuron2}
\end{eqnarray}

In our experiments, we have found that p-bit network with p-bits defined as in Eqs.~(\ref{eq:pbit_synapse}-\ref{eq:pbit_neuron}) converges faster than those defined with Eqs.~(\ref{eq:pbit_synapse2}-\ref{eq:pbit_neuron2}) as shown in Supplementary Figure ~\ref{fig:pbit_comparison} where we emulated the $6\times6$ square-octagonal lattice starting from ordered initial condition.
\begin{figure}[!ht]
	\centering
	\vspace{0pt}
	\includegraphics[width=3.25in,keepaspectratio]{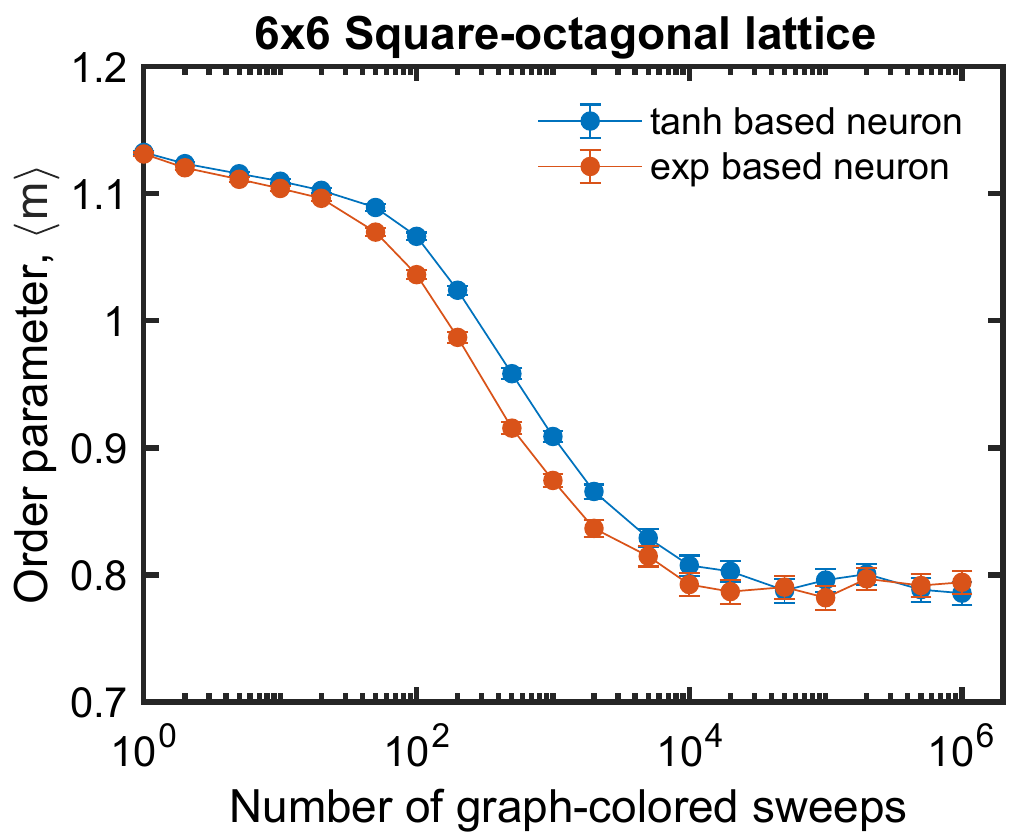}
	\caption{\textbf{Comparison between two types of neuron used for p-bit:} Performance comparison between p-bits described by Eqs.~(\ref{eq:pbit_synapse}-\ref{eq:pbit_neuron}) (red circles) and Eqs.~(\ref{eq:pbit_synapse2}-\ref{eq:pbit_neuron2}) (blue circles). The solid lines serve as visual guide for human eye. When emulating $6\times6$ square-octagonal lattice, we see $\sim2\times$ improvement in the number of sweeps required for convergence (as defined in this work) with the Eqs.~(\ref{eq:pbit_synapse}-\ref{eq:pbit_neuron}).}
	\label{fig:pbit_comparison}
\end{figure}

\begin{figure*}[!ht]
	\centering
	\vspace{0pt}
	\includegraphics[keepaspectratio,width=0.88\columnwidth]{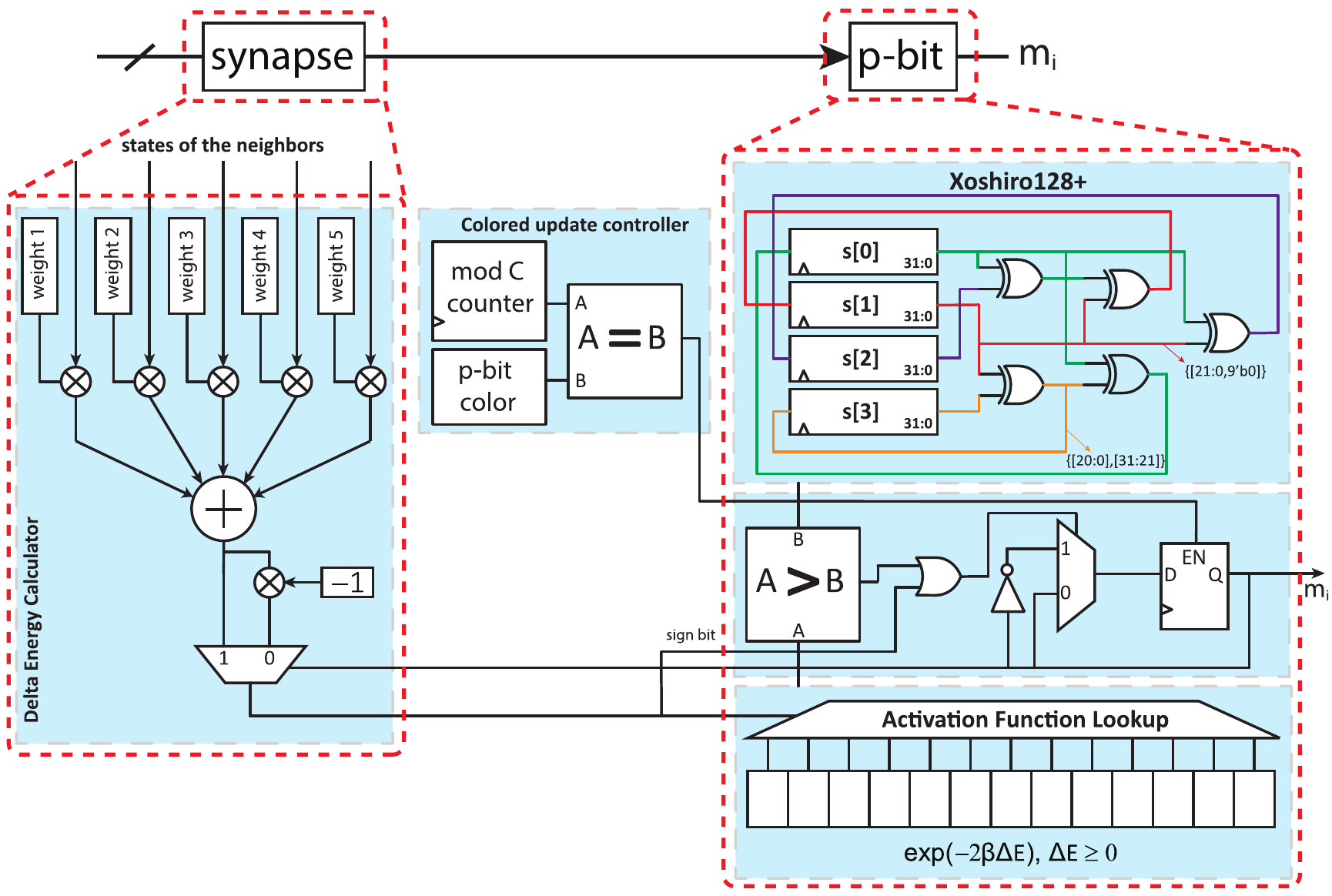}
	\caption{\textbf{Simplified block diagram of a p-bit along with its synapse as implemented on FPGA:} The p-bit as defined in Eq.~(\ref{eq:pbit_neuron}) is modeled with three blocks on the right. The Xoshiro128+ \cite{Blackman2018} pseudorandom number generator on the top-right serves as the source of entropy. The activation function lookup on the bottom right implements the exponential function for negative inputs. For positive inputs, the p-bit is always flipped. The block in the middle right serves as the comparator. The synapse on the left is modeled with an adder consisting of five inputs corresponding to five connections per spin in the replicated p-bit network. The output of the adder is multiplied by the current state of the p-bit and then passed to the activation lookup block of the p-bit. Finally, the colored updating of the p-bits is implemented by pre-assigning each p-bit a color value and adding a local counter with each p-bit. This counter increments at each clock pulse from the system clock and it resets when the maximum color value is reached. When the local counter value becomes equal to the color value of the p-bit, only then the output of the p-bit is passed into the network. For simplicity, we did not show the system clock and other global control signals explicitly.}
	\label{fig:FPGA_Basic_Block_Diagram}
\end{figure*}

\begin{figure}[!ht]
	\centering
	\vspace{0pt}
	\includegraphics[width=3.25in,keepaspectratio]{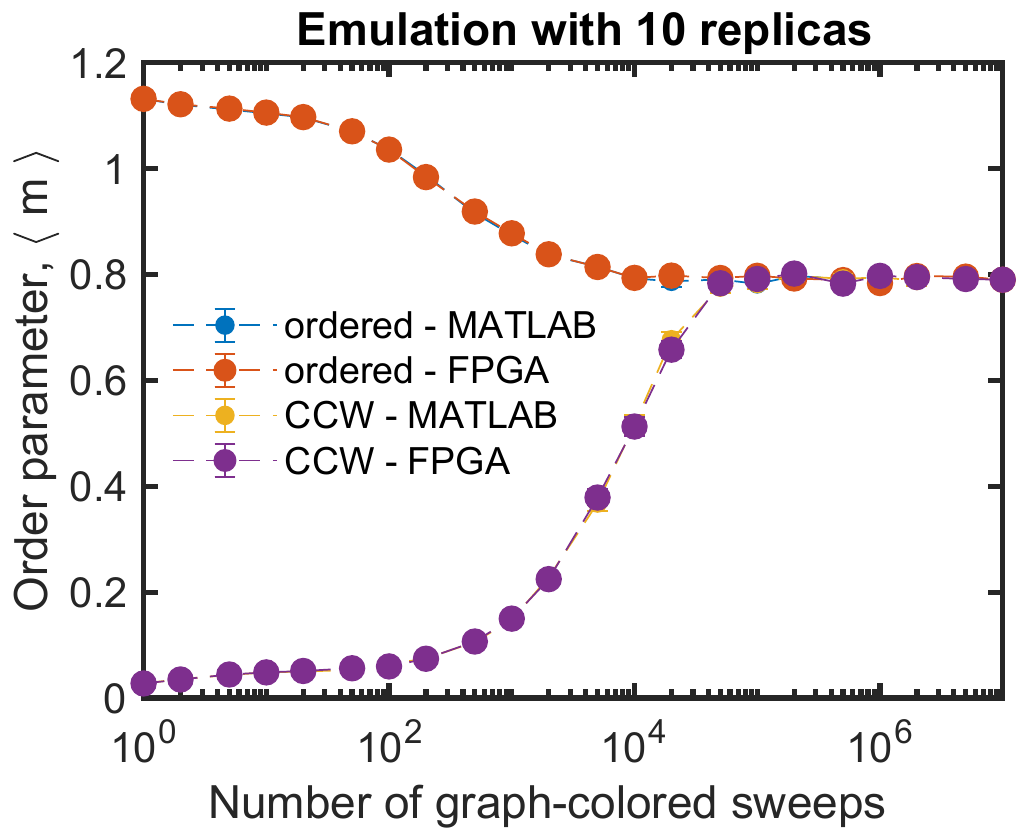}
	\caption{\textbf{Verification of field programmable gate array (FPGA) emulation:} Emulation in FPGA is compared against the simulation in MATLAB to check the validity of the graph-coloring-based FPGA implementation for `ordered' (blue filled circles: MATLAB, red filled circles: FPGA) and `counterclockwise' wound (yellow filled circles: MATLAB, purple filled circles: FPGA) initial conditions. The dashes lines serve as visual guide for human eye. We could implement only the smallest lattice size ($6 \times 6$ which corresponds to 144 qubits and 1440 p-bits) in the FPGA because of the unavailability of resources in FPGA. This plot shows that the FPGA results are in excellent agreement with MATLAB results. This also demonstrates the fact that both ordered and counterclockwise wound initial conditions lead to the same equilibrium value.}
	\label{fig:FPGA_MATLAB_equiv}
\end{figure}

The plot for $6\times6$ lattice in Fig.~3(a) was generated from emulation of a graph-colored p-computer on FPGA (using Virtex\textsuperscript{\textregistered} UltraScale+\textsuperscript{TM} - xcvu9p provided by AWS F1 instances and with a clock frequency of 125 MHz clock corresponding to 8 ns clock period). The p-bit network emulated in AWS FPGA follows a similar local weighted p-bit architecture proposed in \cite{sutton2020autonomous}, although we did not use the autonomous mode. Instead, we used the graph coloring approach and was implemented by making use of local counters assigned with each p-bit which keeps track of the p-bits to be updated based on their assigned color. A basic block diagram of a p-bit implemented on FPGA is shown in Supplementary Figure~\ref{fig:FPGA_Basic_Block_Diagram}. In the qubit lattice, each qubit is connected to at most three other q-bits (except for the qubits at the boundary along the horizontal direction, which are connected to either one or two other qubits). The p-bits in the translated network, on the other hand, are connected to at most five other p-bits because of the replicas above and below. This allows us to design a fast and local synapse where at most only five inputs are to be considered. We use an activation function lookup table to mimic the operation of $\exp{(-2\beta x)}$ function. We used 16-bit weight precision (1 bit for sign and 15 bits for value) along with Xoshiro128+ pseudorandom number generator (to perform accept-reject logic based on Metropolis-Hastings algorithm) which yielded visually similar quality sweeps as obtained from  MATLAB\textsuperscript{\textregistered} (where `mt19937ar' random number generator was used) as evident from Supplementary Figure  \ref{fig:FPGA_MATLAB_equiv}. In Fig.~3(a), only the smallest lattice was actually implemented in real AWS FPGA. Unfortunately, we did not have enough resources (see Supplementary Table ~\ref{tab:DWAVE_tab1} for resources utilized for $6\times6$ lattice) in AWS FPGA to emulate bigger lattice sizes but we expect that given enough resources our p-bit architecture would follow the projections made in Fig.~3(a). From Supplementary Figure  \ref{fig:FPGA_MATLAB_equiv}, we can clearly see that the results from FPGA almost exactly follow the results from MATLAB\textsuperscript{\textregistered}. To summarize, for the next three larger p-computer curves, we project the time to convergence for other data points by (1) simulating the lattices on MATLAB\textsuperscript{\textregistered}, (2) collecting the number of sweeps required to converge (following the same criterion set in \cite{King2021}), and (3) multiplying the number of sweeps from MALTAB\textsuperscript{\textregistered} by 16 ns per sweep.

\begin{table}[!ht]
	\centering
	\caption{FPGA resources utilized in the implementation of a p-computer emulator for $6\times6$ square-octagonal lattice.}
	\vspace{0.5 cm}
	\begin{tabular}{cccccc} \toprule
		&\text{LUT}  & \text{FF}  & \text{BRAM} & \text{URAM} & \text{Power}\footnote{Estimated using Xilinx's power estimation tool}\\
		&\text{usage} & \text{usage} & \text{usage} & \text{usage} & \text{(W)}\\ \midrule
		\text{Absolute} &  678534 & 567698 & 24.01 & 43 & 44.875 \\ [0.2em]
		\text{\%} & 57.45 & 24.01 & 9.19 & 4.48 & \\ \bottomrule
	\end{tabular}
	\label{tab:DWAVE_tab1}
\end{table}

We also note that it is possible to run multiple p-computers in parallel because they are relatively cheaper than quantum annealing processor which needs to consume a large amount of power ($\sim25$ kW) for cooling. Using Xilinx power estimation tool to estimate the power consumed by the FPGA implementation of $6\times6$ square-octagonal lattice (see Table~\ref{tab:DWAVE_tab1}) we get $\sim45$ W for our implementation. Therefore at the same power consumption level 500 p-computer (consuming $\sim22.5$ kW power) could run in parallel giving 500 sweeps at each time point simultaneously yielding a significant improvement in `total' run-time of the problem. With the inclusion of nanomagnet-based p-bits, the power consumption for p-computer would go down even more. This should not be possible with QA processors because of their huge power consumption per machine for which they have to run these parallel runs sequentially.

\section{Justification for using exponential fitting}
\label{app:chapDWAVE_exp_fit}

Initial probability vector, $P_0$ is a $N\times1$ column vector where $N = 2^n$ and $n$ is the number of p-bits. Since, for every independent run, we start from a single state, therefore it has the following form: 
\begin{equation}
	P_{0}=\left[\begin{array}{cccccccc}
		0 & 0 & \ldots  & 0 & 1 & 0 & \ldots  & 0
	\end{array} \right]^{\text{T}}
\end{equation}
where the location of `1' corresponds to counterclockwise wound (or any other fixed) initial state. Let $W$ be our $N\times N$ transition matrix and $\lambda_0,{{\lambda }_{1}},{{\lambda }_{2}},\ldots ,{{\lambda}_{N}}$ are its eigenvalues in the descending order. Then $\lambda_0=1$ and
\begin{equation}
	{{P}_{0}}={{\alpha }_{0}}\left\vert {{\lambda }_{0}} \right\rangle +{{\alpha }_{1}}\left\vert {{\lambda }_{1}} \right\rangle +{{\alpha }_{2}}\left\vert {{\lambda }_{2}} \right\rangle +\ldots +{{\alpha }_{N-1}}\left\vert {{\lambda }_{N-1}} \right\rangle 
\end{equation}
where $\vert\lambda_i\rangle$ is the $N\times 1$ eigenvector corresponding to eigenvalue  $\lambda_i$. After applying $W$ matrix for $k$ times (which corresponds to running the emulator for $k$ sweeps), we get the probability vector
\begin{eqnarray}
	{{P}^{\left(k\right)}}&=&{{W}^{k}}{{P}_{0}} \nonumber\\ 
	&=&{{\alpha }_{0}}\left\vert {{\lambda }_{0}} \right\rangle +{{\alpha }_{1}}{\lambda }_{1}^k\left\vert {{\lambda }_{1}} \right\rangle +{{\alpha }_{2}}{\lambda }_{2}^k\left\vert {{\lambda }_{2}} \right\rangle +\ldots +{{\alpha }_{N}}{\lambda }_{N}^k\left\vert {{\lambda }_{N}} \right\rangle \nonumber\\
	&=&{{\alpha }_{0}}\left\vert {{\lambda }_{0}} \right\rangle +{{\alpha }_{1}}{{e}^{k\ln {{\lambda }_{1}}}}\left\vert {{\lambda }_{1}} \right\rangle +{{\alpha }_{2}}{{e}^{k\ln {{\lambda }_{2}}}}\left\vert {{\lambda }_{2}} \right\rangle +\ldots + {{\alpha }_{N}}{{e}^{k\ln {{\lambda }_{N}}}}\left\vert {{\lambda }_{N}} \right\rangle
\end{eqnarray}

Now, the quantity $m$ we are trying to evaluate has the form:
\begin{eqnarray}
	m &=&{{\beta }_{0}}{{p}_{0}}+{{\beta }_{1}}{{p}_{1}}+\ldots +{{\beta }_{N}}{{p}_{N}} \nonumber\\ 
	&=&{{\beta }_{0}}\left( {{\alpha }_{0}}{{\lambda }_{00}}+{{\alpha }_{1}}{{\lambda }_{01}}{{e}^{k\ln {{\lambda }_{1}}}}+\ldots +{{\alpha }_{N}}{{\lambda }_{0N}}{{e}^{k\ln {{\lambda }_{N}}}} \right) + {{\beta }_{1}}\left( {{\alpha }_{0}}{{\lambda }_{10}}+{{\alpha }_{1}}{{\lambda }_{11}}{{e}^{k\ln {{\lambda }_{1}}}}+\ldots +{{\alpha }_{N}}{{\lambda }_{1N}}{{e}^{k\ln {{\lambda }_{N}}}} \right) \nonumber\\ 
	&+&\ldots \,+ {{\beta }_{N}}\left( {{\alpha }_{0}}{{\lambda }_{N0}}+{{\alpha }_{1}}{{\lambda }_{N1}}{{e}^{k\ln {{\lambda }_{1}}}}+\ldots +{{\alpha }_{N}}{{\lambda }_{NN}}{{e}^{k\ln {{\lambda }_{N}}}} \right)
\end{eqnarray}	

where $\beta_i$ is the corresponding $m$ value for the $i$\textsuperscript{th} state, $p_i$ is the $i$\textsuperscript{th} element of $P^{(k)}$ and $\lambda_{ij}$ is the $i$\textsuperscript{th} element of eigenvector $\vert\lambda_j\rangle$. Rearranging, we get
\begin{eqnarray}	
	m&=&{{\alpha }_{0}}\left( {{\beta }_{0}}{{\lambda }_{00}}+{{\beta }_{1}}{{\lambda }_{10}}+\ldots +{{\beta }_{N}}{{\lambda }_{N0}} \right) + {{\alpha }_{1}}\left( {{\beta }_{0}}{{\lambda }_{01}}+{{\beta }_{1}}{{\lambda }_{11}}+\ldots +{{\beta }_{N}}{{\lambda }_{N1}} \right){{e}^{k\ln {{\lambda }_{1}}}} \nonumber\\ &+& \ldots \,+ {{\alpha }_{N}}\left( {{\beta }_{0}}{{\lambda }_{0N}}+{{\beta }_{1}}{{\lambda }_{1N}}+\ldots +{{\beta }_{N}}{{\lambda }_{NN}}\right){{e}^{k\ln {{\lambda }_{N}}}} \nonumber
\end{eqnarray}

Therefore, we do expect the output of the emulations to be a sum of many exponentials. In our experiments, we found that a fit with a sum of two exponentials matches the data better than just a single exponential fitting.

\begin{figure}[!ht]
	\centering
	\vspace{0pt}
	\includegraphics[width=3.25in,keepaspectratio]{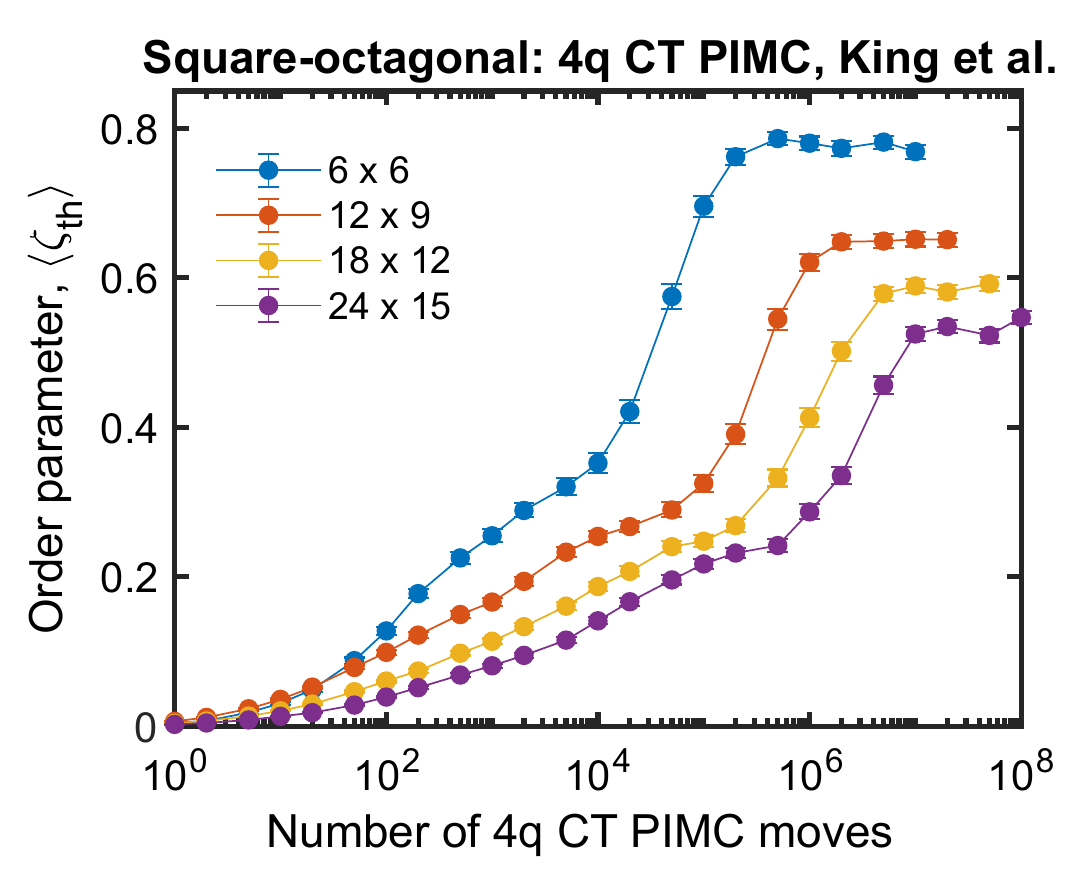}
	\caption{\textbf{Equilibrium value determination:} We run the 4 qubit continuous time path integral Monte Carlo (4q-CT PIMC) code provided by King et al. for all four lattices (blue: $6\times6$, red: $12\times9$, yellow: $18\times12$, purple: $24\times15$) and extract the equilibrium value. Filled circles represent data points while the solid lines serve as visual guide for human eye. We use these equilibrium values to calculate the errors in Fig.~2(c) of the main article.}
	\label{fig:DWAVE_CCW}
\end{figure}

\section{Results from the 4q CT-PIMC algorithm}
\label{app:chapDWAVE_4qCTPIMC}

We show the results of our simulation results with the 4q CT-PIMC algorithm for all four lattice sizes in Supplementary Figure ~\ref{fig:DWAVE_CCW}. Here we have defined one 4q-CT PIMC move as a sample. We fit these curves as described before and determine the equilibrium value. We use these results to compare the p-computer results in Fig.~2(c).

\section{Effect of the quality of random number generator}
\label{app:chapDWAVE_RNG_quality}

\begin{figure}[!ht]
	\centering
	\vspace{0pt}
	\includegraphics[width=3.25in,keepaspectratio]{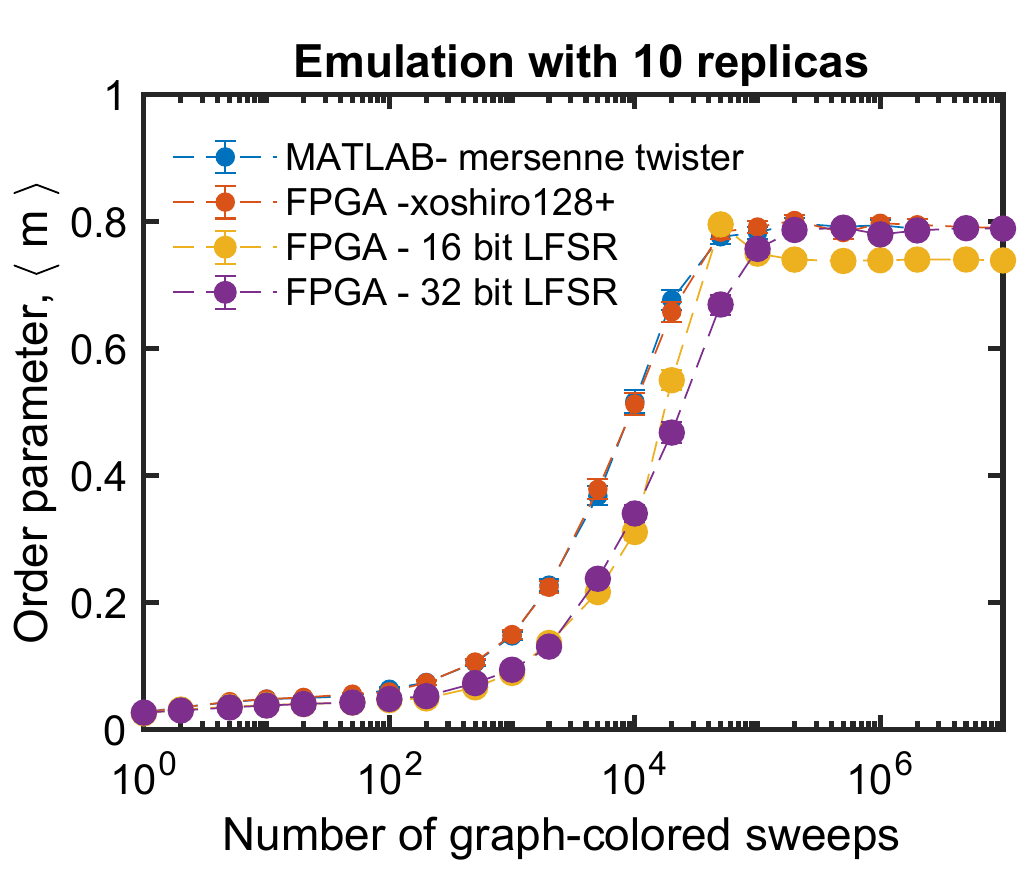}
	\caption{\textbf{Effect of the quality of random numbers:} The quality of random numbers is important for this benchmarking problem. The use of 16-bit linear feedback shift regrister (LFSR) per p-bit (yellow filled circles) does not provide the correct saturation value. 32-bit LFSR (purple filled circles) per p-bit seems to reach the correct saturation value but takes longer to converge. The use of costly but superior quality hardware RNG such as xoshiro128+ (red filled circles) provides excellent performance when compared with Mersenne-twister (blue filled circles) implemented on MATLAB. Stochastic magnetic tunnel junction based compact p-bits can be useful for providing high-quality random numbers with high throughput.}
	\label{fig:rngQ}
\end{figure}

\vspace{0.5cm}

In this supplementary note, we also present our study of the effect of the quality of random number generators for the benchmarking problem. Supplementary Figure ~\ref{fig:rngQ}, illustrates the importance of using high-quality random numbers. A cheap random number generator such as a linear feedback shift register (LFSR) is not good enough. In software programs, one can generate very good random numbers using high-quality RNG such as Mersenne twister but it requires many clock cycles to generate one random number. In hardware, one can also generate moderately good quality random numbers (as shown in Supplementary Figure ~\ref{fig:rngQ}, xoshiro128+ works very well for this problem), but one needs to use too much digital footprint per RNG. The nano-magnet-based `compact' p-bits can provide a solution to both problems: it can generate true random numbers at a very high speed and requires very less hardware footprint.

\section{Scaling with p-bits}
\label{app:chapDWAVE_p_scaling}
\begin{figure}[!ht]
	\centering
	\vspace{0pt}
	\includegraphics[width=3.25in,keepaspectratio]{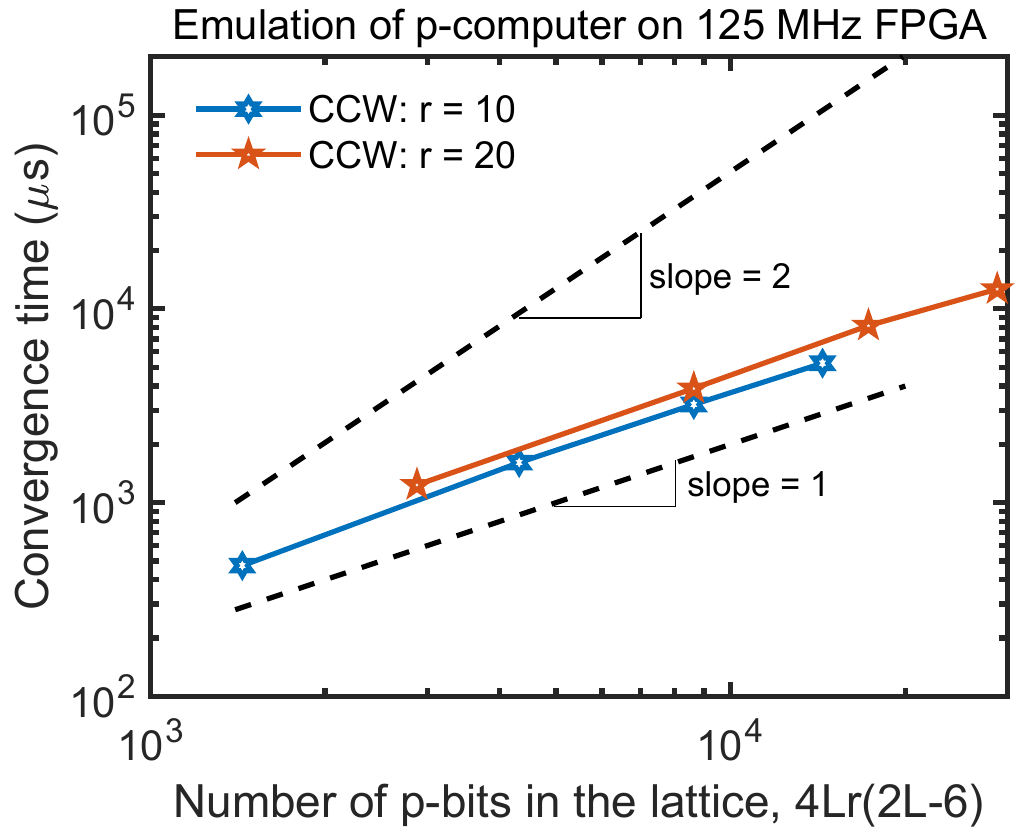}
	\caption{\textbf{Convergence time scaling with p-bits:} The scaling of convergence time with the number of p-bits in the lattice is shown for counterclockwise (CCW) wound initial conditions. As can be seen here, the curves are almost linear yielding an approximately $rN_Q$ scaling with the number of p-bits (red stars: 20 replicas, blue stars: 10 replicas). The small gap between two different replica sets can be attributed to their slight different saturation values. For better visual comparison, curves with slope = 1 and  slope = 2 (dashed lines) are also shown.}  
	\label{fig:NrScaling}
\end{figure}

Finally, we note that the scaling of convergence time with the number of p-bits approximately $ rN_Q$ for this problem when $r$ is sufficiently large such that trotterization error is small. This has been shown in Supplementary Figure ~\ref{fig:NrScaling}(a). The small differences in the curves for 10 and 20 replicas arise because they saturate at slightly different values (i.e., saturation values obtained using 10 replicas are more erroneous than saturation values obtained using 20 replicas). We also note that the error with the 4q CTPIMC reduces a little compared to 10 replicas when we use 20 replicas.

\section{More details on the slopes of the convergence time plot}
\label{app:chapDWAVE_TTS_Slopes}

The slopes of the fitted curves are extracted using MATLAB's curve fitting tool with `Levenberg-Merquardt' algorithm and with `Bisquare' robustness. They are presented in Supplementary Table~\ref{tab:DWAVE_tab2}:

\begin{table}[!ht]
	\centering
	\caption{Slopes of the fitted curves in Fig.~5.}
	\vspace{0.5 cm}
	\begin{tabular}{lr@{}} \toprule
		\text{Algorithm/Hardware}  & \text{Slope}  \\ \midrule
		GC MATLAB (CPU) & 1.946 \\ 
		Optimized 4q CTPIMC (CPU) & 1.575 \\ 
		125 MHz p-computer (FPGA) & 1.047 \\ 
		QA Processor & 0.441 \\ 
		[0.2em]\bottomrule
	\end{tabular}
	\label{tab:DWAVE_tab2}
\end{table}

\pagebreak

\end{document}